\documentclass{IEEEtran}

\usepackage[export]{adjustbox}
\usepackage{booktabs}
\usepackage{float}
\usepackage{flushend}
\usepackage{graphicx}
\usepackage{ifpdf}
\usepackage[utf8]{inputenc}
\usepackage{listings}
\usepackage{multirow}
\usepackage[caption=false,font=footnotesize]{subfig}
\usepackage{url}
\usepackage{tabularx}
\usepackage{textcomp}
\usepackage{xcolor}
\usepackage{xspace}
\usepackage{cite}

\definecolor{listinggray}{gray}{0.95}
\definecolor{darkgray}{gray}{0.7}
\definecolor{commentgreen}{rgb}{0, 0.4, 0}
\definecolor{darkblue}{rgb}{0, 0, 0.6}
\definecolor{purple}{rgb}{0.6, 0, 0.6}
\definecolor{middleblue}{rgb}{0, 0, 0.75}
\definecolor{darkred}{rgb}{0.4, 0, 0}
\definecolor{brown}{rgb}{0.5, 0.5, 0}
\definecolor{dkgreen}{rgb}{0,0.5,0}
\definecolor{orange}{rgb}{1,.5,0}
\definecolor{dandelion}{cmyk}{0,0.29,0.84,0}

\usepackage[normalem]{ulem}
\makeatletter
\def\cyanuwave{\bgroup \markoverwith{\lower3.5\p@\hbox{\sixly \textcolor{cyan}{\char58}}}\ULon}
\def\reduwave{\bgroup \markoverwith{\lower3.5\p@\hbox{\sixly \textcolor{red}{\char58}}}\ULon}
\def\blueuwave{\bgroup \markoverwith{\lower3.5\p@\hbox{\sixly \textcolor{blue}{\char58}}}\ULon}
\font\sixly=lasy6 
\makeatother

\def\BibTeX{{\rm B\kern-.05em{\sc i\kern-.025em b}\kern-.08em
    T\kern-.1667em\lower.7ex\hbox{E}\kern-.125emX}}

\newif\ifdraft{}

\ifdraft{}
  \newcommand{\amnote}[1]{ \textcolor{blue} { ***andrem: #1 }}
  \newcommand{\jhanote}[1]{ {\textcolor{red} { ***shantenu: #1 }}}
  \newcommand{\mtnote}[1]{ {\textcolor{orange} { ***matteo: #1 }}}
\else
  \newcommand{\amnote}[1]{}
  \newcommand{\jhanote}[1]{}
  \newcommand{\mtnote}[1]{}
\fi

\newcommand{\B}[1]{\textbf{#1}\xspace}

\newcommand{\up}{\vspace*{-0.5em}}

\newcommand{\mr}[1]{\multirow{2}{*}{#1}}%

\lstdefinestyle{myListing}{
  frame=single,
  backgroundcolor=\color{listinggray},
  language=C,
  basicstyle=\ttfamily \footnotesize,
  breakautoindent=true,
  breaklines=true
  tabsize=2,
  captionpos=b,
  aboveskip=0em,
  belowskip=-2em,
}

\lstdefinestyle{myPythonListing}{
  frame=single,
  backgroundcolor=\color{listinggray},
  language=Python,
  basicstyle=\ttfamily \footnotesize,
  breakautoindent=true,
  breaklines=true
  tabsize=2,
  captionpos=b,
}

\ifpdf{}
  \DeclareGraphicsExtensions{.pdf, .jpg, .tif}
\else
  \DeclareGraphicsExtensions{.eps, .jpg, .ps}
\fi

\begin{document}

\title{Design and Performance Characterization of RADICAL-Pilot on Leadership-class Platforms}

\author{Andre Merzky$^{1*}$, Matteo Turilli$^{1*}$, Mikhail Titov$^{1}$, Aymen Al-Saadi$^{1}$, Shantenu Jha$^{1}$$^{,2}$\\
   \small{\emph{$^{1}$ Rutgers, the State University of New Jersey, Piscataway, NJ 08854, USA}}\\
   \small{\emph{$^{2}$ Brookhaven National Laboratory, Upton, NY 11973, USA}} \\
   \small{\emph{$^{*}$ Joint First Authors}}\\
   \up
}

\maketitle

\begin{abstract}
Many extreme scale scientific applications have workloads comprised of a large
number of individual high-performance tasks. The Pilot abstraction decouples
workload specification, resource management, and task execution via job
placeholders and late-binding. As such, suitable implementations of the Pilot
abstraction can support the collective execution of large number of tasks on
supercomputers. We introduce RADICAL-Pilot (RP) as a portable, modular and
extensible pilot-enabled runtime system. We describe RP's design, architecture
and implementation. We characterize its performance and show its ability to
scalably execute workloads comprised of tens of thousands heterogeneous tasks on
DOE and NSF leadership-class HPC platforms. Specifically, we investigate RP's
weak/strong scaling with CPU/GPU, single/multi core, (non)MPI tasks and Python
functions when using most of ORNL Summit and TACC Frontera. RADICAL-Pilot can be
used stand-alone, as well as the runtime for third-party workflow systems.
\end{abstract}


\section{Introduction}\label{sec:intro}

An analysis of workloads and
applications~\cite{antypas2013nersc,liu2020characterization} on pre-exascale
platforms suggests that scientific workloads increasingly require multiple
heterogeneous tasks, instead of a monolithic single task. This trend was
confirmed by the 2020 ACM Gordon Bell Special Prize for High Performance
Computing-Based COVID-19 Research, where all four
finalists~\cite{casalino2021ai} involved sophisticated workflows.

Even as HPC simulations increasingly become important generators of data for
powerful and expensive ML models, ML/AI components are substituting
traditional HPC sub-components~\cite{jia2020pushing}, and innovative methods
coupling ML components to steer HPC simulations are
emerging~\cite{casalino2021ai}. Thus, workflows with diverse components (e.g.,
physics-based simulations, data generation and analysis, and ML/AI tasks) will
become increasingly common on extreme-scale platforms. Such workflows will
encompass high-throughput function calls, ensembles of MPI-based simulations,
and AI-driven HPC simulations. There are no ``turnkey solutions'' to support
diverse tasks across multiple heterogeneous platforms, with the necessary
performance, scale and required throughput. As workflows emerge as an
important development paradigm for extreme-scale applications, the role and
importance of runtime systems to support the resource management and execution
requirements~\cite{hwang2016resource} of concurrent heterogeneous tasks will
increase.

Pilot systems~\cite{turilli2018comprehensive} address two apparently
contradictory requirements: accessing HPC resources via their centralized
schedulers, and letting applications independently schedule tasks on the
acquired portion of resources. By implementing multi-level scheduling and
late-binding, pilot systems lower the overhead of task scheduling, enable higher
task execution throughput, and allow greater control over the resources acquired
to execute workloads. As such, pilot systems provide a promising starting point
for the resource management and execution requirements of concurrent
heterogeneous tasks.

Traditionally, pilot systems were used to enable high-throughput task execution
on HPC platforms~\cite{chiu2010pilot}. Pilot systems now  implement both
pilot and runtime capabilities to serve a much wider phase space of use
cases~\cite{deelman2018future}. Specifically, pilot systems must support the
effective and efficient execution of single/multi core/GPU/node tasks,
implemented either as executables or functions, on diverse HPC platforms, with
heterogeneous hardware and execution environments. In fact, a computational task
is a generalized term, usually indicating either a stand-alone process with
input, output, termination criteria, and dedicated resources; or a function
executed in a dedicated environment. A task can be used to represent an
independent simulation or data processing analysis, running on one or more nodes
of a HPC machine, may require MPI or OpenMP but, often, may be executed within a
single compute node. Further, pilot systems need to meet the unprecedented
requirements of upcoming exascale computing, supporting dynamic partitioning of
resources, adaptive task scheduling policies and diverse placement and launching
methods.

In response to the aforementioned requirements, we introduce RADICAL-Pilot
(RP)~\cite{merzky2018using}, a Pilot-enabled runtime system that implements the
pilot paradigm as outlined in Ref.~\cite{turilli2018comprehensive}, alongside
advanced runtime placement and launching capabilities. RP is implemented in
Python and provides a well-defined API and usage modes.  RP serves as a runtime
system for workflow management
systems~\cite{dakka2018high,oleynik2017high,turilli2016analysis}, and it has
been integrated with EnsembleToolkit, Parsl, Swift/T, PanDA and QCFractal. More
in general, RP is designed as a building block~\cite{turilli2019middleware} that
can be integrated with any workflow management system implementing the task
abstraction, e.g., Pegasus, BeeFlow or Taverna. Further, RP pilot and runtime
capabilities are independent and can also be integrated with third-party systems
like, for example, the Flux runtime system~\cite{ahn2014flux}. Once integrated,
RP provides pilot capabilities to Flux's scheduler and task-launching
mechanisms.

This paper has two main contributions: (i) a detailed description of the design
and architecture of RP, with an analysis of RP unique features and capabilities;
and (ii) a detailed analysis of RP's scaling performance when executing
workloads comprised of homogeneous and heterogeneous tasks, implemented as
executables or functions, on leadership class platforms. Together, those two
contributions allow to uncover the overheads of specific RP components and
illustrate how they were avoided in order to optimize overall scale and
performance. Specifically, we characterize RP weak and strong scaling
performances on most of the resources available on Titan, Frontera and Summit,
using up to 392,000 cores and 24,582 GPUs to execute 24,552 heterogeneous
executable tasks and $126\times10^6$ Python function tasks.

RP works on multiple HPC platforms. We focus our experiments on open
academic research machines \textemdash{} Titan, Summit and Frontera, that
offered and still offer the highest degree of concurrent execution in the open science community.
We configured RP to overcome existing bottlenecks, so that both the
performance and scalability of RP are determined by system software limits.
Specifically, we show that the launch rate of tasks is dominated by
overheads arising from the use of the OpenMPI Runtime Environment (ORTE) and
PMIx Reference RunTime Environment (PRRTE), or by the file system performance
of the HPC platforms. The results of our experiments support the idea that
partitioning resources at pilot-level will enable better scaling on the
upcoming exascale platforms.

Although RP is a vehicle for research in scalable computing, it also supports
production-grade science. Currently, RP is used by applications from diverse
domains, including high-energy physics, earth and climate sciences, biomolecular
sciences and drug discovery. Since 2018, RP has been used to support more than
$10^7$ node-hours on DoE (Andes, Titan, Rhea, Summit, Lassen, Theta),
NSF (Blue Waters, Frontera and XSEDE Stampede, Stampede2, SuperMIC, Comet,
Bridges), and European (Archer and SuperMUC) HPC platforms. RP has been the
core runtime system for eight DoE INCITE awards and one NSF PRAC award. It has
also served as the workhorse for DoE's National Virtual Biotechnology Laboratory
COVID19 drug discovery pipeline~\cite{lee2021scalable}, collectively consuming a
further estimated lower bound of $10^7$ node-hours on several of the DoE
and NSF HPC machines listed above.

In~\S\ref{sec:related}, we discuss existing pilot systems and highlight the
distinctive capabilities of RP. \S\ref{sec:arch} discusses the design and
architecture of RP and~\S\ref{sec:exp} describes the core experiments and
results of the paper. Overall, the contributions of this paper show the benefits
and limitations of using the pilot abstraction and architectural pattern for
executing applications with heterogeneous tasks on HPC platforms, including
leadership-class machines. Further, our analysis and results
clarify the role that pilot systems will play in the upcoming exascale
supercomputers.

\section{Related Work}\label{sec:related}

Runtime systems support the execution of units of work on computing resources.
Specifically, runtime systems can be designed to operate at different levels of
a software stack. In this paper, we focus on a type of runtime system that sits
above the operating system and can manage the execution of both executable and
function tasks.

Charm++~\cite{kale1993charm++}, HPX~\cite{kaiser2014hpx} and
Cilk~\cite{blumofe1995cilk} are runtime systems that enable scalable multi-task
execution but assume vertical and dedicated programming models, depending on
specific compilers and/or application programming interfaces (APIs).
Flux~\cite{ahn2014flux} is an example of a more general-purpose runtime system
that supports scalable execution of executable tasks on HPC platforms. Flux
supports task scheduling, placement and execution. RADIAL-Pilot belongs to the
same class of runtime systems as Flux but focuses on the efficient management of
heterogeneous tasks and HPC resources via pilots.

Many scientific workloads have heterogeneous
tasks~\cite{preto2014fast,cheatham2015impact} that can benefit from being
executed at scale on leadership-class HPC platforms. Nonetheless, a tension
exists between these workloads' requirements and HPC systems capabilities as,
traditionally, HPC systems are designed to best support monolithic workloads.
Several software systems address this tension, but their adoption presents
limitations, including type of workloads and resources supported, how resources
are selected and acquired, the scale at which workloads can be executed, the
programming paradigm they support, and the lack of development and maintenance.

Since 1995, more than twenty pilot systems have been
developed~\cite{turilli2018comprehensive}. Most of these systems are tailored to
specific use cases, workloads, resources, interfaces or development models. Some
notable examples are: (i) HTCondor with Glidein on OSG~\cite{pordes2007open}, a
widely used pilot system for the execution of mostly single-core tasks; (ii)
the pilot systems developed for the LHC communities (e.g.,
PanDA~\cite{maeno2014evolution}, GlideinWMS~\cite{sfiligoi2008glideinwms},
DIRAC~\cite{casajus2010dirac} and CernVM Co-Pilot~\cite{harutyunyan2012cernvm})
which execute millions of jobs a week
and are specialized in supporting Large Hadron Collider (LHC) workloads on
specific platforms like the Worldwide LHC Computing Grid and the CERN
cloud; (iii) Falkon~\cite{raicu2007falkon}, specifically designed
to support function-level parallelism as opposed to process-level parallelism;
(iv) FireWorks~\cite{jain2015fireworks}, designed to support function-level
parallelism and small-scale process-level parallelism on HPC resources; and (v)
GWpilot~\cite{rubio2015gwpilot} that enables the use of arbitrary scheduling
algorithms with the GridWay meta-scheduler, and supports a limited number of
non-MPI use cases.

Several workflow management systems use pilots to support the execution of
multi-task applications on HPC machines. For example,
Parsl~\cite{babuji2019parsl} high-throughput executor provides pilot job
capabilities on HPC and cloud platforms but with limited MPI support.
Pegasus~\cite{deelman2015pegasus} uses Glidein and providers like
Corral~\cite{deelman2005pegasus}, Makeflow~\cite{albrecht2012makeflow} and
FireWorks~\cite{jain2015fireworks} to enable users to manually start pilots on
HPC resources via master/worker frameworks like Queue~\cite{bui2011work} or
FireLauncher~\cite{jain2015fireworks}. Swift~\cite{zhao2007swift} can use
Falkon~\cite{raicu2007falkon} or the Coasters~\cite{hategan2011coasters} pilot
system, with or without JETS~\cite{wozniak2013jets}, to support MPI and non-MPI
jobs on HPC and cloud platforms, but requires an application-level
domain-specific language.

Diverse tools enable the execution of multi-task workloads on HPC machines,
using job arrays and leveraging MPI either as a launch method or as a container
for multiple tasks. All of them reach limited scale or require low-level
programming for multi-task applications. For example, PBS Job
Arrays~\cite{simmerman2013eden} enable concurrent execution of multiple
instances of the same executable within a single job submission. The Application
Level Placement Scheduler (ALPS)~\cite{karo2006application} enables the
concurrent execution of a limited number of different executables on CRAY
systems. CRAM~\cite{gyllenhaal2014enabling} parallelizes the execution of
multiple executables by statically bundling them into a single MPI executable.
TaskFarmer~\cite{taskfarmer} and Wraprun~\cite{wraprun} enable single-core or
single-node executables to be run within a single \texttt{mpirun} and
\texttt{aprun} allocation.

\up
\section{Design of RADICAL-Pilot (RP)}\label{sec:arch}

RADICAL-Pilot (RP) is a pilot system designed to address the main limitations of
the tools described in~\S\ref{sec:related}, either by implementing missing
capabilities or by enabling integration among independent software systems. RP
addresses research challenges related to efficiency, effectiveness, scalability
and both workload and resource heterogeneity. RP requires managing the flow of
information across multiple components, distributed across different machines.
Further, RP has to enable scheduling, placement and launching of heterogeneous
tasks on heterogeneous resources, with minimal overheads and maximal resource
utilization.

Accordingly, RP enables the execution of one or more workloads comprised of
heterogeneous tasks on one or more HPC platforms. Tasks can be implemented as
stand-alone executables, free functions or class methods. These tasks can be
placed, launched and executed on CPUs, GPUs and other accelerators, on the same
pilot or across multiple pilots. As a pilot system, RP schedules tasks
concurrently and sequentially, depending on available resources, and defines
scheduling policies for executing tasks on the acquired resources.

RP offers five unique features when compared to other pilot systems that execute
workloads on HPC platforms: (1) concurrent execution of tasks with five types of
heterogeneity; (2) concurrent execution of multiple workloads on a single pilot,
across multiple pilots and across multiple HPC platforms; (3) support of all
major HPC batch systems to acquire and manage computing resources; (4) support
of fifteen methods to launch tasks; and (5) integration with third-party
workflow and runtime systems. The five types of task heterogeneity supported by
RP are: (1) type of task (executable, function or method); (2) parallelism
(scalar, MPI, OpenMP, or multi-process/thread); (3) compute support (CPU and
GPU); (4) size (1 hardware thread to 8000 compute nodes); and duration (zero
seconds to 48 hours).

Every pilot system requires scheduling a job on an HPC machine via its batch
system to acquire resources, which makes supporting diverse platforms with the
same code base challenging. RP uses RADICAL-SAGA~\cite{merzky2015saga} to
support all the major batch systems: \texttt{Slurm}, \texttt{PBSPro},
\texttt{Torque}, \texttt{LGI}, \texttt{Cobalt}, \texttt{LSF} and
\texttt{LoadLeveler}. Further, as a runtime system, RP supports the following
methods to perform task placement and launching: \texttt{aprun} and
\texttt{ccmrun/mpirun\_ccmrun} on Cray; \texttt{jsrun},
\texttt{dplace/mpirun\_dplace}, \texttt{runjob} and \texttt{POE} on IBM;
\texttt{srun} on Slurm; \texttt{ibrun} on TACC; and \texttt{ORTE},
\texttt{PRRTE}, \texttt{orte\_lib}, \texttt{ssh}, \texttt{rsh}, \texttt{mpirun},
\texttt{mpiexec}, \texttt{mpirun\_mpt}, \texttt{mpirun\_rsh} and \texttt{fork}
on multiple platforms.

Supporting the concurrent execution of heterogeneous tasks via different batch
systems and diverse placing/launching methods requires specific design features.
Particularly challenging is to enable extensibility and scalability within a
single system, avoiding fragmentation into multiple special-purpose systems. RP
is designed to enable localized changes to the existing code base to add new
capabilities required by tasks, and new platforms to acquire resources. Further,
RP can instantiate multiple instances of its components, distributing them
across available resources, depending on the platform specifics. Each component
can be individually configured so as to enable further tailoring while
minimizing code refactoring.

RP improves capabilities already available in other pilot systems by not adding
any software requirement on the HPC platforms and by exposing an API specific to
the pilot abstraction. RP does not require the deployment of services and
daemons, nor to access any dedicated interface or port on the login nodes of the
HPC platforms. Instead, RP uses capability already available like \texttt{ssh},
\texttt{gsissh} or \texttt{scp}. RP API enables the development of tools on top
of the pilot abstraction, cleanly separating resource selection, acquisition and
scheduling from task definition, scheduling, placement and execution. RP API is
implemented in Python, avoiding the need for a domain-specific language.

The need to support both task and resource-level heterogeneity while avoiding
the development of independent special-purpose systems, imposes design
trade-offs. RP's configurability allows it to perform well for diverse resources
and workloads, but RP is not optimized for any specific use case. Our
configuration-based approach is powerful but it can require extensive tailoring,
especially for scenarios other than those supported by default. Further, the
dependence on the software environment of each HPC platform makes deployment
fragile as every change in the environment may require changes in RP's
configuration. This is mitigated by a dedicated integration testing framework
but remains a main challenge of RP's maintainability and portability. Porting RP
to a new platform may require just a new configuration file or writing a
connector for a batch system not yet supported or an executor for a new (MPI)
launching system. While developing connectors and executors requires system
programming skills, they are standalone components that require no changes to
the rest of RP code base.

\up
\subsection{Architecture and Implementation}

\begin{figure}
  \centering
  \includegraphics[trim=0 3 0 1,clip,width=0.49\textwidth]{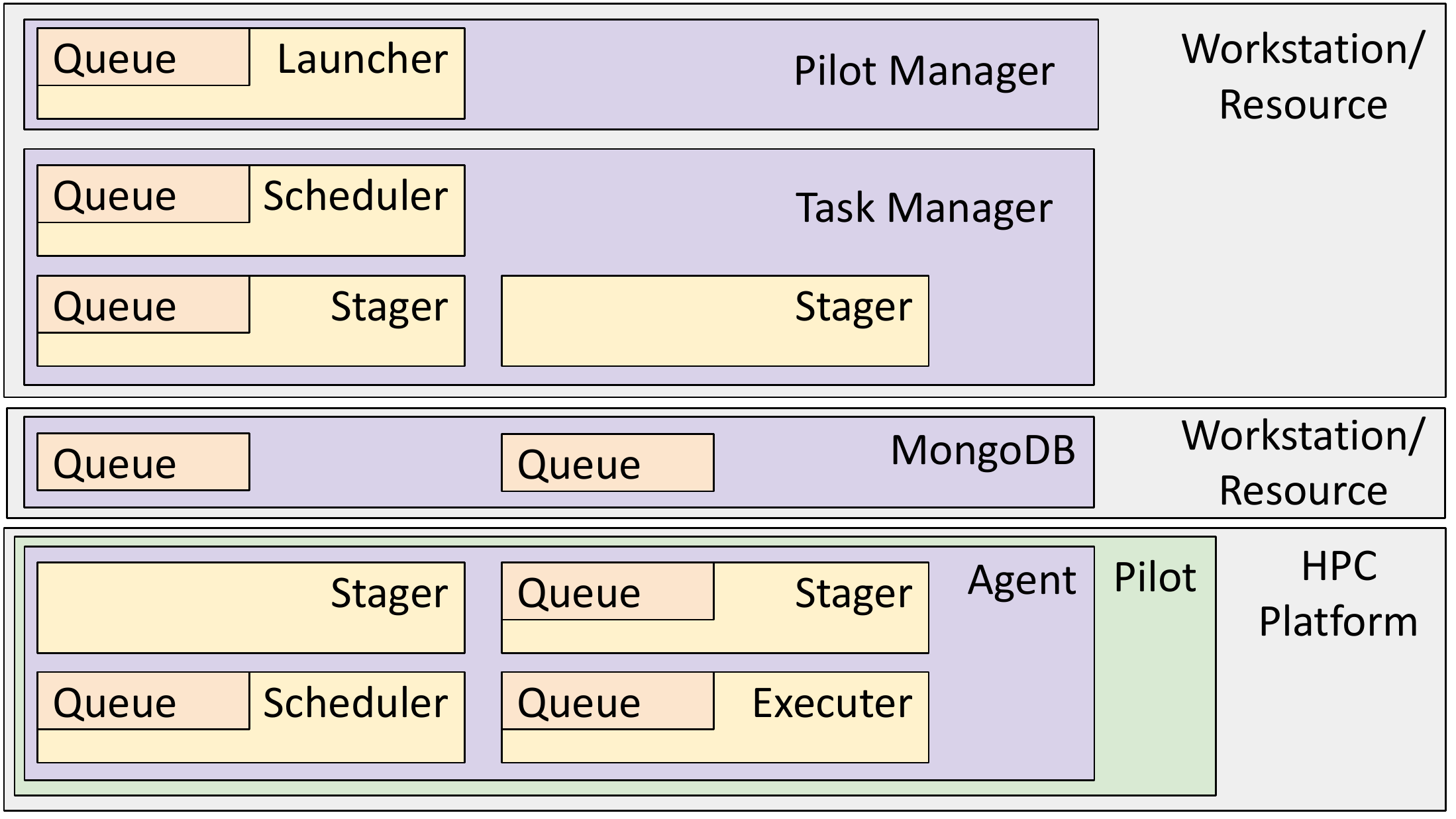}
  \up\up\up
  \caption{RADICAL-Pilot architecture.}\label{fig:arch-overview}
  \up\up\up
\end{figure}

RP implements two main abstractions: Pilot and Task. Pilots are placeholders for
computing resources, where resources are represented independent from
architectural details. Tasks are units of work, specified either as an
application executable, function or method, alongside resource and execution
environment requirements. Currently, RP implements executors for Python
functions but executors for other languages can be added without requiring
changes in RP architecture.

RP offers an API to describe both pilots and tasks, alongside classes and
methods to manage acquisition of resources, scheduling of tasks on those
resources, and the staging of input and output files. Reporting capabilities
update the user about ongoing executions and tracing capabilities enable
post-mortem analysis of workload and runtime behavior.

Architecturally, RP is a distributed system with four modules: PilotManager,
TaskManager, Agent and DB (Fig.~\ref{fig:arch-overview}, purple boxes). Modules
can execute locally or remotely, communicating and coordinating over TCP/IP\@,
and enabling multiple deployment scenarios. For example, users can run the
PilotManager and TaskManager locally, and distribute the DB and one or more
instances of the Agent on remote HPC infrastructures. Alternatively, users can
run all RP modules on a local or on a remote resource.

PilotManager, TaskManager and Agent have multiple components where some are used
only in specific deployment scenarios, depending on both workload requirements
and resource capabilities. Some components can be instantiated concurrently to
enable RP to manage multiple pilots and tasks simultaneously. This allows to
scale throughput and enables component-level fault tolerance. Components are
coordinated via a dedicated ZeroMQ-based communication mesh, which introduces
runtime and infrastructure-specific overheads, but improves overall scalability
of the system and lowers component complexity. ZeroMQ was chosen over other
messaging systems for its superior performance, but also for its communication
patterns Publish/Subscriber and Router/Dealer that match our use case very well.
Components can have different implementations, and configuration files can
tailor RP to specific resources types, workloads or scaling requirements.

PilotManager has a main component called `Launcher'
(Fig.~\ref{fig:arch-overview}). The Launcher uses resource configuration files
to define the number, placement and properties of the Agent's components of each
Pilot. Currently, configuration files are made available for the major USA NSF
and DOE production HPC resources, but users can provide new files or alter
existing configuration parameters at runtime, both for a single and multiple
pilots. This enables supporting of campus-level clusters (e.g., Traverse at
Princeton University or Amarel at Rutgers University) and lab-level private
clusters.

Agent has four main components: two Stagers (one for input and one for output
data), Scheduler and Executor (Fig.~\ref{fig:arch-overview}). Multiple instances
of the Stager and Executor components can coexist in a single Agent. Depending
on the architecture of the target HPC platform, the Agent's components can
individually be placed on login nodes, MOM nodes, compute nodes or any other
combination. ZeroMQ communication bridges connect the Agent components, creating
a network to support the transitions of the tasks through components.

Once instantiated, each Agent's Scheduler gathers information from the resource
manager, retrieving the number of cores/GPUs held by the pilot on which the
Agent is running and the partitioning of cores/GPUs across nodes. Depending on
requirements, the Agent's Scheduler assigns cores and GPUs from one or more
nodes to each task. For example, cores on a single node are assigned to
multithreaded tasks, while cores on topologically close nodes are assigned to
MPI tasks to minimize communication overheads. Three scheduling algorithms are
currently supported: ``Continuous'' for nodes organized as a continuum,
``Torus'' for nodes organized in a n-dimensional torus, as found, for example,
on IBM BG/Q, and ``Tagged'' to pin the execution of tasks on specific nodes.

The Agent's Scheduler passes the tasks on to one of the Agent's Executors, which
use resource configuration parameters to derive the placement and launching
command of each task. Once the launching command is determined, depending on the
task parameters and characteristics of the execution environment, the Executors
execute those commands to spawn the application processes. Two spawning
mechanisms are available: \texttt{Popen} (based on Python) and \texttt{Shell}
(based on \texttt{/bin/sh}). Executors collect task exit codes and communicate
the freed cores to the Scheduler.

\up\up
\subsection{Execution Model}\label{sub:modcomp}

\begin{figure}
  \centering
  \includegraphics[trim=0 0 0 0,clip,width=0.49\textwidth]{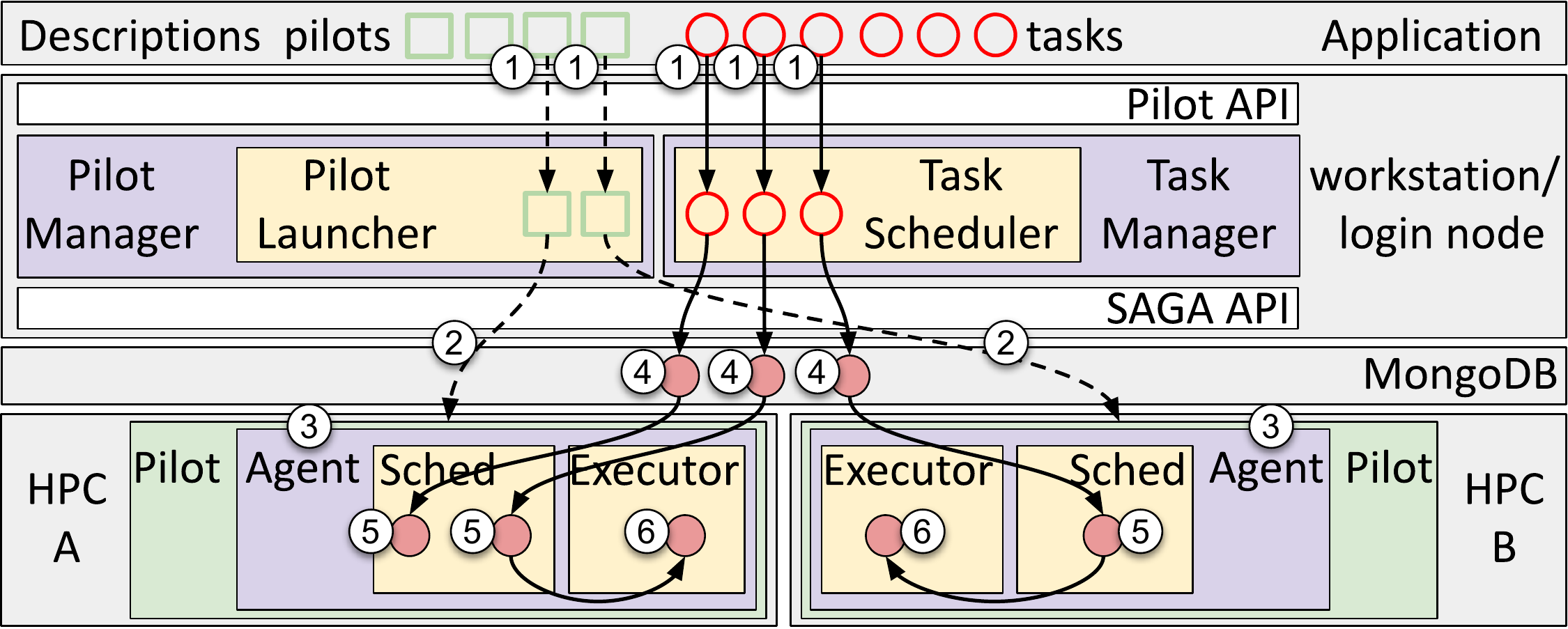}
  \up\up\up
  \caption{RADIAL-Pilot execution model.}\label{fig:exec-model}\up\up\up
\end{figure}

Pilots and tasks are described via the Pilot API and passed to the RP runtime
system (Fig.~\ref{fig:exec-model}, 1). The PilotManager submits pilots on one or
more resources via the SAGA API (Fig.~\ref{fig:exec-model}, 2). The SAGA API
implements an adapter for each supported resource type, exposing uniform methods
for job and data management. Once a pilot becomes active on a resource, it
bootstraps the Agent module (Fig.~\ref{fig:exec-model}, 3).

The TaskManager schedules each task to an Agent (Fig.~\ref{fig:exec-model}, 4)
via a queue on a MongoDB instance. This instance is used as the RP DB module to
communicate task descriptions between the TaskManager(s) and the Agent(s). Each
Agent pulls tasks from the DB module (Fig.~\ref{fig:exec-model}, 5) and
schedules (Fig.~\ref{fig:exec-model}, 6) each task on an Executor upon resource
availability (e.g., number of cores or GPUs). The Executor sets up the task's
execution environment and then spawns the task for execution.

Once a task returns from its execution, the Executor communicates to the
Scheduler that resources have been freed and the scheduling loop can proceed.
Once the workload has been executed, the runtime system is terminated to reduce
resource utilization. Multiple workloads can be described and executed within
the time boundaries of resource availability.

When required, the input data of a task are either pushed or pulled by the
Agent, depending on data locality and sharing requirements. Similarly, the
output data of a task are staged out by the Agent and TaskManager to a specified
destination, e.g., a filesystem accessible by the Agent or the user workstation.
Both input and output staging are optional, depending on task requirements. The
actual file transfers are enacted via RADICAL-SAGA, and currently support
(gsi)-scp, (gsi)-sftp, Globus Online and local filesystem operations.

\up\up
\subsection{Extensibility and Integration}\label{sub:integration}

\begin{figure*}[]
  \up\up
  \centering
  \subfloat[][]{
    \includegraphics[width=0.31\textwidth]{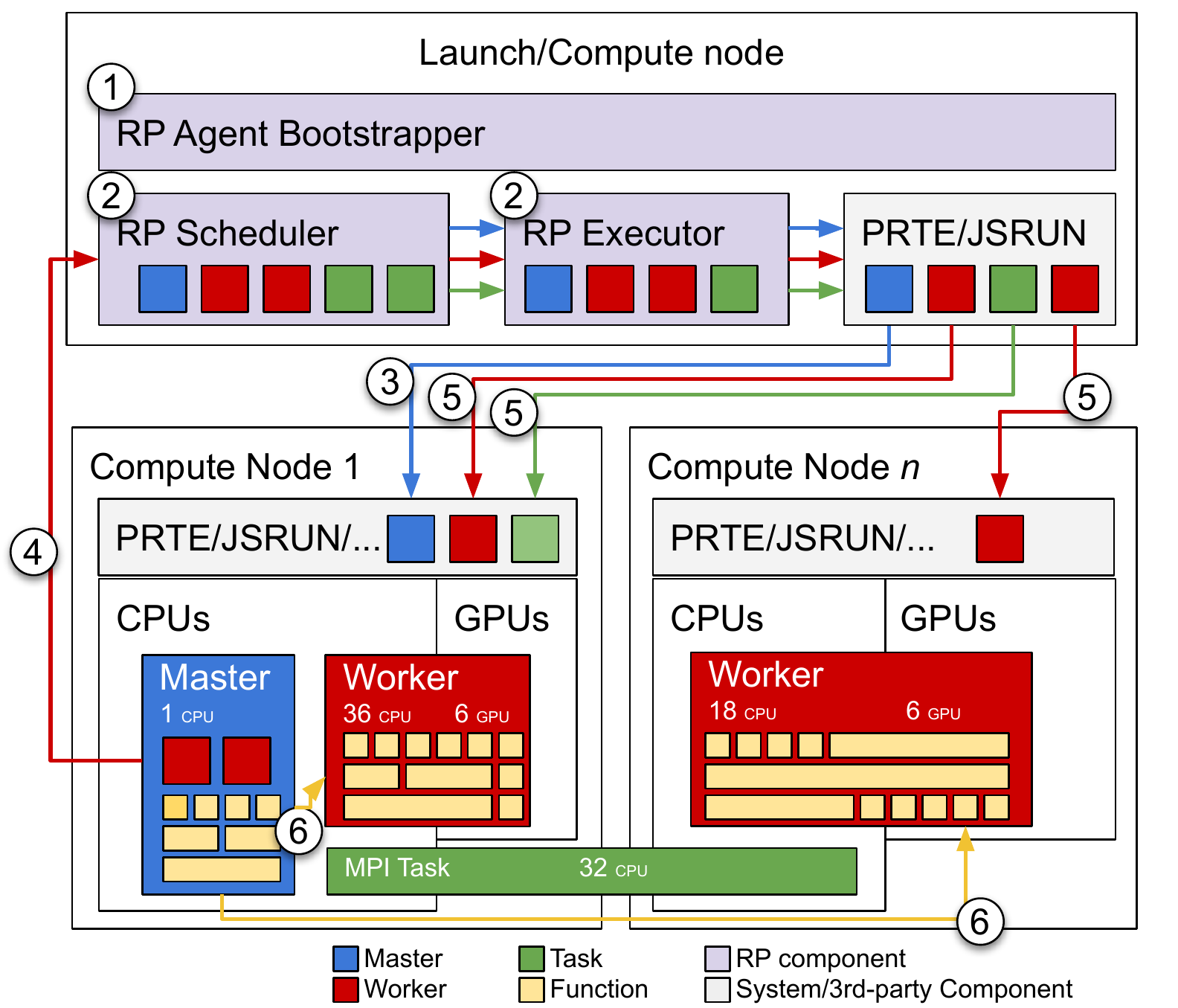}
    \label{sfig:raptor}}
  \hfill
  \subfloat[][]{
    \includegraphics[width=0.34\textwidth]{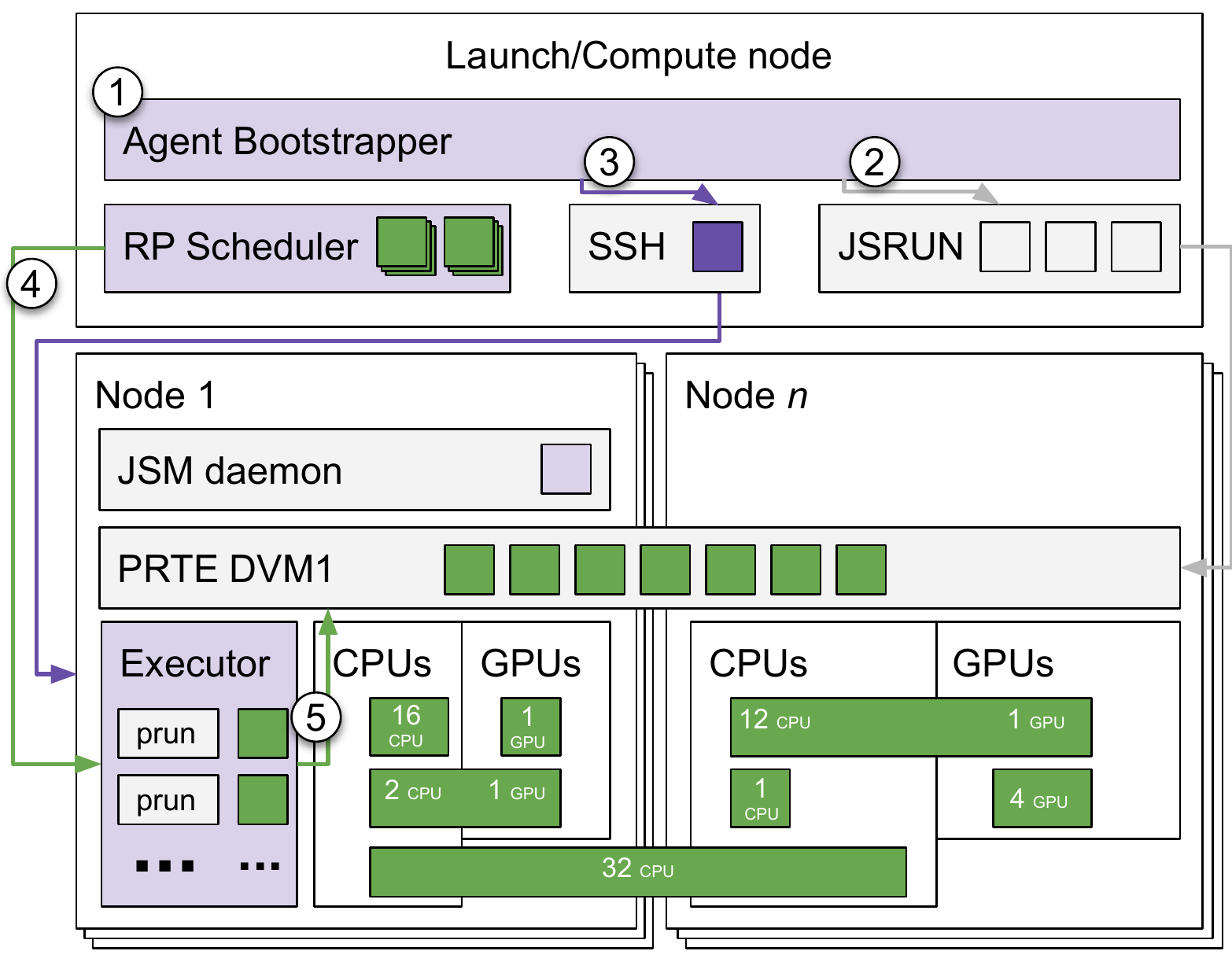}
    \label{sfig:multidvm}}
  \hfill
  \subfloat[][]{
    \includegraphics[width=0.24\textwidth]{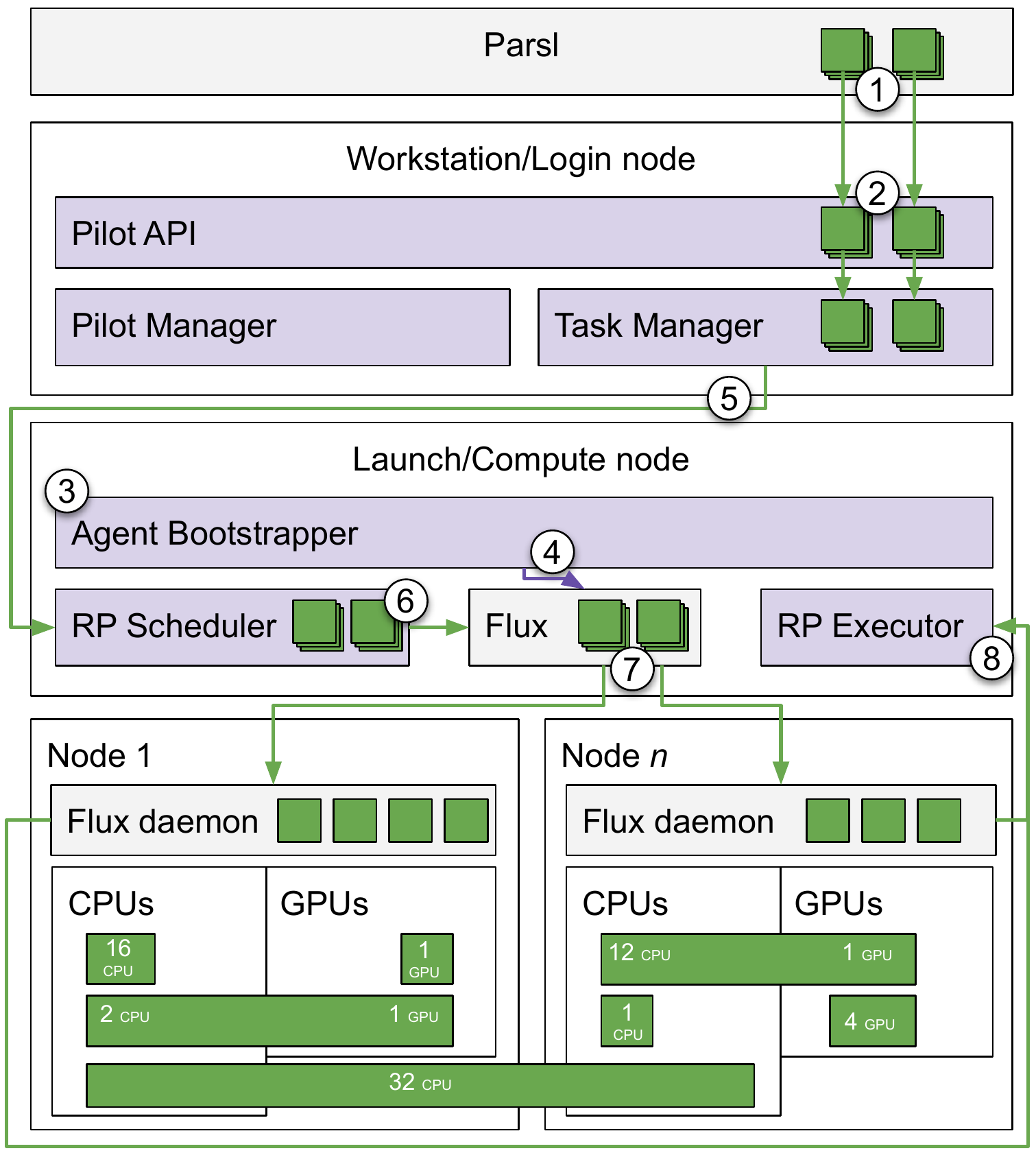}
    \label{sfig:flux}}
  \up
  \caption{Pilot-based task execution frameworks implemented using RADIAL-Pilot.
    Numbers indicates the process of task execution. Blue box = RAPTOR master;
    red box = RAPTOR worker; green box = tasks; purple box = RP component; gray
    box = third-party software component. (a) RAPTOR's masters/workers are
    special type of tasks executed via the standard RP capabilities. (b) Each
    DVM spans multiple compute nodes and one RP Executor is used for each DVM to
    execute tasks on those compute nodes. (c) Integration with both user-facing
    (Parsl) and resource-facing (Flux) software systems, does not alter RP
    execution model: task are described in Parsl, scheduled by RP and placed and
    launched by Flux.}
    \up\up\up
  \label{fig:lms}
\end{figure*}

The design, configurability and execution model of RP enable architectural and
behavioral customizations alongside the integration of RP with third-party
software systems. Fig.~\ref{fig:lms} illustrates three paradigmatic examples:
(1) Fig.~\ref{sfig:raptor} shows the design of a master/worker framework called
RAPTOR built with RP to support effective and efficient execution of Python
functions and single-node tasks at scale; (2) Fig.~\ref{sfig:multidvm} shows the
use of multiple PRRTE Distributed Virtual Machines (DVMs) to partition the
concurrent execution of heterogeneous tasks at scale; and (3)
Fig.~\ref{sfig:flux} shows how RP enables integration with third-party software
systems, either by coding just a new launch method (Flux) or a dedicated
connector for RP API (ParSL).

RP's execution model supports the execution of arbitrary tasks, including
specialized tasks which can hook into RP's communication protocols. That
mechanism has been used to implement RAPTOR (Fig.~\ref{sfig:raptor}): first one
or more master tasks are scheduled, placed and launched, followed by one worker
task per compute node. Once both have successfully bootstrapped, each master
directly coordinates its pool of workers to schedule and execute the specified
Python function calls or tasks. Assuming functions and tasks that require at
most a single compute node, RAPTOR enables unprecedented scaling and performance
on leadership-class HPC platforms.

Specific capabilities can be implemented in an Agent component, without
modifying the overall execution model of RP. For example, we extended an Agent's
Executor to support multiple PRRTE DVMs (Fig.~\ref{sfig:multidvm}). Available
resources are partitioned across the DVMs and the Executor places tasks across
available DVMs. Currently, tasks can be placed round-robin or by tagging each
task to a specific DVM.

Finally, RP execution model is amenable to integration with third-party software
that implement functionalities needed by RP. For example, in the integration
with Flux (Fig.~\ref{sfig:flux}), the Agent's Staging\_in component queues tasks
to the Flux's scheduler that, in turn, places and launches those tasks across
the resources held by RP's Agent.

\up\up
\subsection{Programmability, Tracing and Profiling}\label{sub:profiling}

RP exposes an API with 5 classes: Session, PilotManager, PilotDescription,
TaskManager, TaskDescription~\cite{rp_api_url}. Users use those classes and
their methods to describe resources, pilots and tasks; create managers for both
resources and tasks, and then launch the execution of the workload. The
application waits for the workload to complete before returning control, making
RP well suited for stand-alone applications as opposed to those which require
interactive programming~\cite{rp_examples_url}). The API is implemented in pure
Python and users import RP as a module in their Python applications.

The distributed, modular, and concurrent design of RP introduces complexities
with both usability and performance overheads. We developed a tracer to enable
postmortem performance analysis, collecting up to 200 unique events across RP
components, and a profiling library called RADICAL-Analytics (RA). RA
synchronizes traces' timestamps and enables time series analysis that we use to
characterize RP's performance. The tracer adds some overhead, included in the
results presented in this paper. By using buffered I/O and small data structures
we can keep that overhead manageable. For example, a typical run of experiment 1
in~\S\ref{sec:exp} lasts \(1045.5 \pm 29.4s\) without tracing and \(1069.2 \pm
49.5s\) with tracing. Tracing thus increases the runtime of about \(2.5\%\), and
also slightly increases the noise of the measurements.

\up\up
\section{Performance Characterization}\label{sec:exp}

We characterize the performance of RP with homogeneous and heterogeneous
workloads, executing emulated, synthetic and real-world tasks implemented both
as executables and Python function calls. We characterize the scaling and
performance of RP in terms of mean time to execution (TTX) of the workload,
compute resource utilization (RU), and RP Agent's runtime overheads (OVH).

\up\up
\subsection{Experiments Design}\label{ssec:exp_design}

As seen in~\S\ref{sec:arch}, Figs.~\ref{fig:arch-overview}
and~\ref{fig:exec-model}, RP reduces every workload to the execution of a set of
tasks on its Agent. The Agent retrieves tasks individually or in bulk and
executes them on the previously acquired HPC resources. The execution of
workloads requires the interplay of all RP components and their supporting
infrastructure.  As such, the characterization of TTX, RU and OVH depends on how
each Agent component performs.

As explained in~\S\ref{sec:intro} and~\S\ref{sec:arch}, the Pilot abstraction
and RP Agent enable the execution of tasks both concurrently and sequentially.
Above a certain number of tasks, the workload cannot be executed with full
concurrency, even on the largest HPC platforms currently available. In this
situation, sequential ``batched'' execution incurs overheads determined by the
systems and resources used to manage the execution.

Our experiments are designed to measure the overhead that the Agent, third-party
systems, and the HPC platform add to the execution of the workload. Overhead
captures the time spent \textit{not} executing tasks while the resources were
available to RP. This overhead determines a partial utilization of the available
computing time for executing the workload and, therefore, a certain degree of
inefficiency of its execution. We investigate its growth with increasing number
of tasks and cores.

We designed five experiments to characterize the Agent performance when
executing homogeneous and heterogeneous workloads. Experiment 1 measures the
weak scaling of the Agent by maintaining a constant ratio of homogeneous tasks
to resources. Experiment 2 measures the strong scaling by fixing the number of
homogeneous tasks while varying the amount of resources. Experiments 3 and 4
also measure the weak and strong scaling of the Agent but for heterogeneous
tasks, using multiple DVMs (\S\ref{sec:arch}) and improved scheduling algorithms
to reach higher scale and better performance. Experiment 5 measures the
performance of RP when using RAPTOR (\S\ref{sec:arch}) and a production
workload. Together, experiments 1--5 characterize the performance of RP for
diverse workloads, on diverse HPC platforms and at the largest scales that can
be currently reached on HPC resources available to scientific research.

\begin{figure}
   \centering
   \includegraphics[trim=0 0 0 0,clip,width=0.49\textwidth]{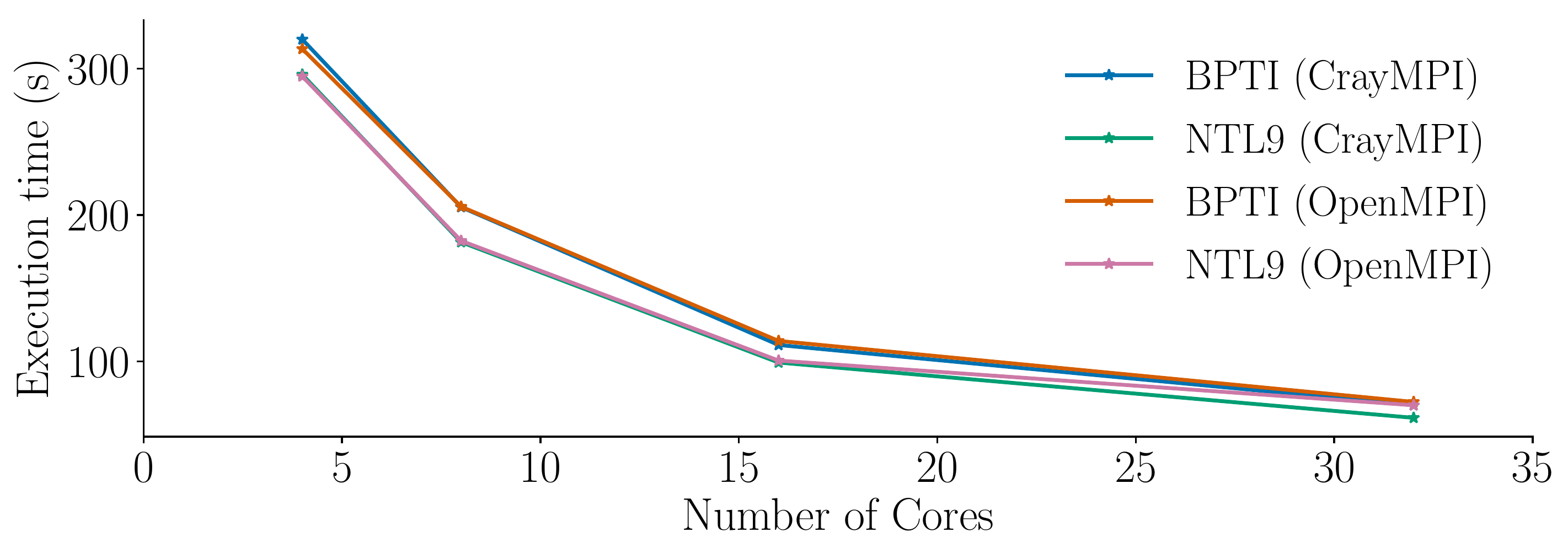}
    \up\up\up\up
   \caption{BPTI, NTL9 scaling on Titan.}\label{fig:bpti-scaling}
    \up\up\up
\end{figure}

Experiments 1 and 2 execute a workload comprised of executable tasks simulating
the molecular dynamics of the bovine pancreatic trypsin inhibitor (BPTI), a
globular protein of 20,521 atoms when fully solvated. Fig.~\ref{fig:bpti-scaling}
shows the scaling behavior of GROMACS for workloads simulating BPTI and another
protein NTL9 with 14,100 atoms to confirm the general scaling behavior.
Although the simulations of both proteins scale sublinearly after 8 cores, 32
cores offer the best relative performance, as measured by execution time. With
larger systems, scaling each task up to 64 cores can become optimal.

MD simulations with multiple GROMACS tasks executed on HPC machines can
experience large performance fluctuations over the runtime. Such
fluctuations would make the separation of RP overheads from resource
fluctuations and runtime variations of the application's tasks difficult, if not
impossible. Thus, we profiled and emulated GROMACS simulations with
Synapse~\cite{merzky2018synapse}. Synapse profiles the compute, memory and I/O
use of an executable and emulates them. Synapse reproduces the computing
activities of the profiled executable, faithfully approximating its time to
completion and resource utilization.

Synapse offers our experiments several advantages over the direct use of the
executable it emulates: (1) simplified and self-contained deployment without
third parties libraries and compilers dependences; (2) high-fidelity replication
of compute patterns of the emulated executables; (3) profiling capabilities
independent of third-parties applications; (4) control over the number of FLOPs
executed; and (5) selective emulation of the type of profiled resources. As
such, Synapse allows greater control, while simplifying deployment and data
analysis without loss of generality of results.

\begin{figure}
  \centering
  \includegraphics[trim=0 0 0 0,clip,width=0.49\textwidth]{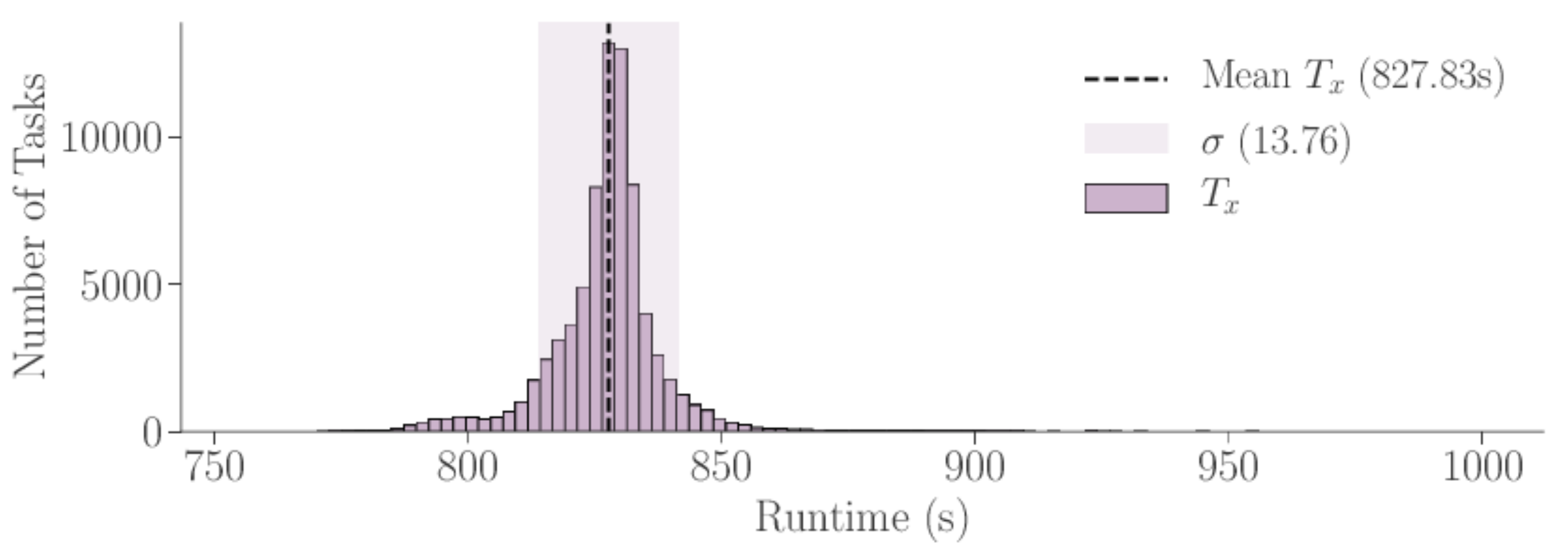}
  \up\up\up\up\up
  \caption{Distribution of the TTX for Synapse emulation of BPTI.
  \label{fig:tx_distribution}}
  \up\up\up
\end{figure}

We emulated the execution of a single GROMACS instance, simulating BPTI for
$\sim$250ps, the baseline in several studies. In this way, we controlled
the runtime noise inherent to executing multiple instances of the same
executable: we measured only the variance of Titan and the predictable variance
of Synapse. Further, we did not emulate I/O activities as the performance
fluctuations of Titan's network file systems would have dominated our
experimental results. Fig.~\ref{fig:tx_distribution} shows the narrow
distribution of Synapse emulations' runtime: the mean is 828s with a standard
deviation of \(\pm\)14s.

Experiments 3 and 4 execute a synthetic workload in which an executable can be
configured to run for an arbitrary amount of time and on an arbitrary number of
cores and/or GPUs, using MPI when spanning multiple compute nodes. In this way,
we can characterize the weak and strong scaling of the Agent when concurrently
executing tasks with four types of heterogeneity: amount and type of parallelism
(scalar, MPI and multi-process/thread); type of required compute support; size;
and duration. Together, these types of heterogeneity represent the requirements
of the diverse use cases supported by RP and offer a worst case scenario for its
performance analysis. Heterogeneity stresses the Agent Scheduler and Executor
components more than homogeneous workloads or workloads with lesser types of
heterogeneity.

Experiment 5 executes a production workload which simulates
the docking of diverse ligands to a protein receptor. The experiment performs
docking of $126\times10^6$ molecules to the 3CLPro\_6LU7\_A\_1\_F receptor,
using OpenEye Python function calls. This workload is a core stage of the DOE
NVBL drug discovery pipeline~\cite{lee2021scalable} to find known drug molecules
that can bind to the severe acute respiratory syndrome coronavirus 2
(SARS-CoV-2). Currently, to our knowledge, RP executes docking calculations at
the largest scales, and a throughput rate that is twice that of highest
published rate~\cite{vermaas2020supercomputing}.

\begin{table*}
   \caption{Experiments setup and results. Weak and strong scaling of
    RADICAL-Pilot for homogeneous tasks (experiments 1--2), heterogeneous tasks
    with multiple DVMs (experiments 3--4), and peak performance of RP and
    RAPTOR (experiments 5).}
   \label{tab:experiments}
   \centering
   \resizebox{\textwidth}{!}{%
   \begin{tabular}{p{0.2cm}          
                   p{1.0cm}          
                   p{1.7cm}          
                   p{1.5cm}          
                   p{1.5cm}          
                   p{1.0cm}          
                   p{1.0cm}          
                   p{2.0cm}          
                   p{2.0cm}          
                   p{1.0cm}          
                   p{1.2cm}          
                   p{1.0cm}          
                   p{1.0cm}          
               }
   \toprule
    \B{ID}                            &  
    \B{HPC      \newline Platform}    &  
    \B{\#Tasks}                       &  
    \B{\#Generations}                 &  
    \B{Task     \newline Runtime (s)} &  
    \B{\#Cores/ \newline Task}        &  
    \B{\#GPUs/  \newline Task}        &  
    \B{\#Cores/ \newline Pilot}       &  
    \B{\#GPUs/  \newline Pilot}       &  
    \B{\#Runs}                        &  
    \B{Scaling}                       &  
    \B{OVH      \newline (\%)}        &  
    \B{RU       \newline (\%)}        \\ 
   \midrule
   \B{1}                             &  
   Titan                             &  
   $2^{5..12}$                       &  
   1                                 &  
   \mr{828\(\pm\)14}                 &  
   \mr{32}                           &  
   \mr{-}                            &  
   $2^{10..17}$                      &  
   \mr{-}                            &  
   8                                 &  
   Weak                              &  
   9..26                             &  
   81..34                            \\ 
   \B{2}                             &  
   Titan                             &  
   $2^{14}$                          &  
   $2^{5..3}$                        &  
                                     &  
                                     &  
                                     &  
   $2^{14-16}$                       &  
                                     &  
   3                                 &  
   Strong                            &  
   9..5                              &  
   85..93                            \\ 
   \B{3}                             &  
   Summit                            &  
   \({3098; 12{,}276}\)              &  
   1                                 &  
   600..900                          &  
   \mr{1..42}                        &  
   \mr{0; 6}                         &  
   \mr{\(43{,}008; 172{,}074\)}      &  
   \mr{\(6144; 24{,}582\)}           &  
   \mr{2}                            &  
   Weak                              &  
   7;9                               &  
   77;41                             \\ 
   \B{4}                             &  
   Summit                            &  
   \({24{,}552; 24{,}784}\)          &  
   $\sim$8; 2                        &  
   500..600                          &  
                                     &  
                                     &  
                                     &  
                                     &  
                                     &  
   Strong                            &  
   2;8                               &  
   76;38                             \\ 
   \B{5}                             &  
   Frontera                          &  
   $126\times10^6$                   &  
   $\sim$300                         &  
   1..120                            &  
   1                                 &  
   -                                 &  
   \(392{,}000\)                     &  
   -                                 &  
   1                                 &  
   -                                 &  
   8                                 &  
   80..98                            \\ 
   \bottomrule
   \end{tabular}
   }
    \up\up
\end{table*}

Table~\ref{tab:experiments} shows the parameters of the five experiments.
Experiment 1 consists of 8 runs designed to measure the weak scaling of RP Agent
with the chosen workload on Titan. Each run executes between 32 and 4096
32-cores tasks on a single pilot with between 1024 and 131,072 cores. The ratio
between the number of tasks executed and the amount of resources acquired is
constant across the 8 runs of the experiment. All the tasks are thus executed
concurrently in a single so-called `generation', i.e., a single set of
concurrent executions. As all the tasks have analogous overheads and all the
tasks execute concurrently, the median of the ideal total execution time (TTX)
of all the tasks should be analogous for all the 8 runs.

Experiment 2 has 3 runs which measure the strong scaling of RP Agent with the
chosen workload on Titan. The ratio between number of tasks and number of cores
of the pilot is the only difference with experiment 1: each run executes 16,384
tasks on a single pilot with between 16,384 and 65,536 cores. Because of the
disparity between the number of cores required by the tasks and the number of
pilot cores, the workload is executed on multiple generations, between 32 and 8.

Experiments 3 and 4 measure how RP Agent scales on Summit, the largest HPC
machine currently available in the USA. We execute between 3098 and 24,784
tasks---heterogeneous for size, duration, and type of parallelism and compute
support---on between 1024 and 4097 of the 4608 compute nodes available on
Summit. Each compute node has 42 CPU cores and 6 GPUs, fully utilized and
partitioned across our workload. For these experiments, we measure resource
utilization (RU) and RP overheads (OVH). In presence of multiple
heterogeneities, the ideal TTX of the workload depends on considering optimal
scheduling policies. RP does not attempt to realize scheduling optimality as
that would depend on the specifics of each workload and resource. Instead, RP
balances the various performance trade offs so as to improve resource
utilization across a variety of workloads and resources. Thus, RP privileges
generality over optimality.

Experiments 3 and 4 also pose a feasibility challenge. Executing at the scale of
near full Summit requires large amount of resource allocations that, in turn,
might not be available on a production, leadership-class machine. Thus, we
reduced the number of runs to two per experiment: a baseline run with a 1/4 of
the total available compute nodes, and a run on almost the whole machine. Those
runs, including the necessary testing and repeated runs for statistical
confidence, consumed around 10,000 node hours, i.e., a full director
discretional allocation on Summit. Thus, we also limited the duration of the
tasks to between 500s and 900s, reducing the pilot job walltime and thus
resource allocation usage to the viable minimum.

Experiment 5 characterizes RP performance when executing 126,471,524 Python
function calls via RAPTOR on 7000 of the 8008 available compute nodes of
Frontera, the largest HPC platform offered by NSF, for a total of 392,000 CPU
cores. For the experiment, we used 70 masters and 6930 workers, i.e., 99 workers
for each master. As we used the Texascale Days at TACC, we execute experiment 5
without incurring allocation limitations of experiments 3 and 4.

\up\up
\subsection{Experiments 1--2: Weak and Strong Scaling with Homogeneous Workloads
on Homogeneous Resources}\label{ssec:exp_ws_ss}

\begin{figure}
  \up\up
  \centering
  \includegraphics[trim=0 0 0 80,clip,width=0.49\textwidth,valign=t]{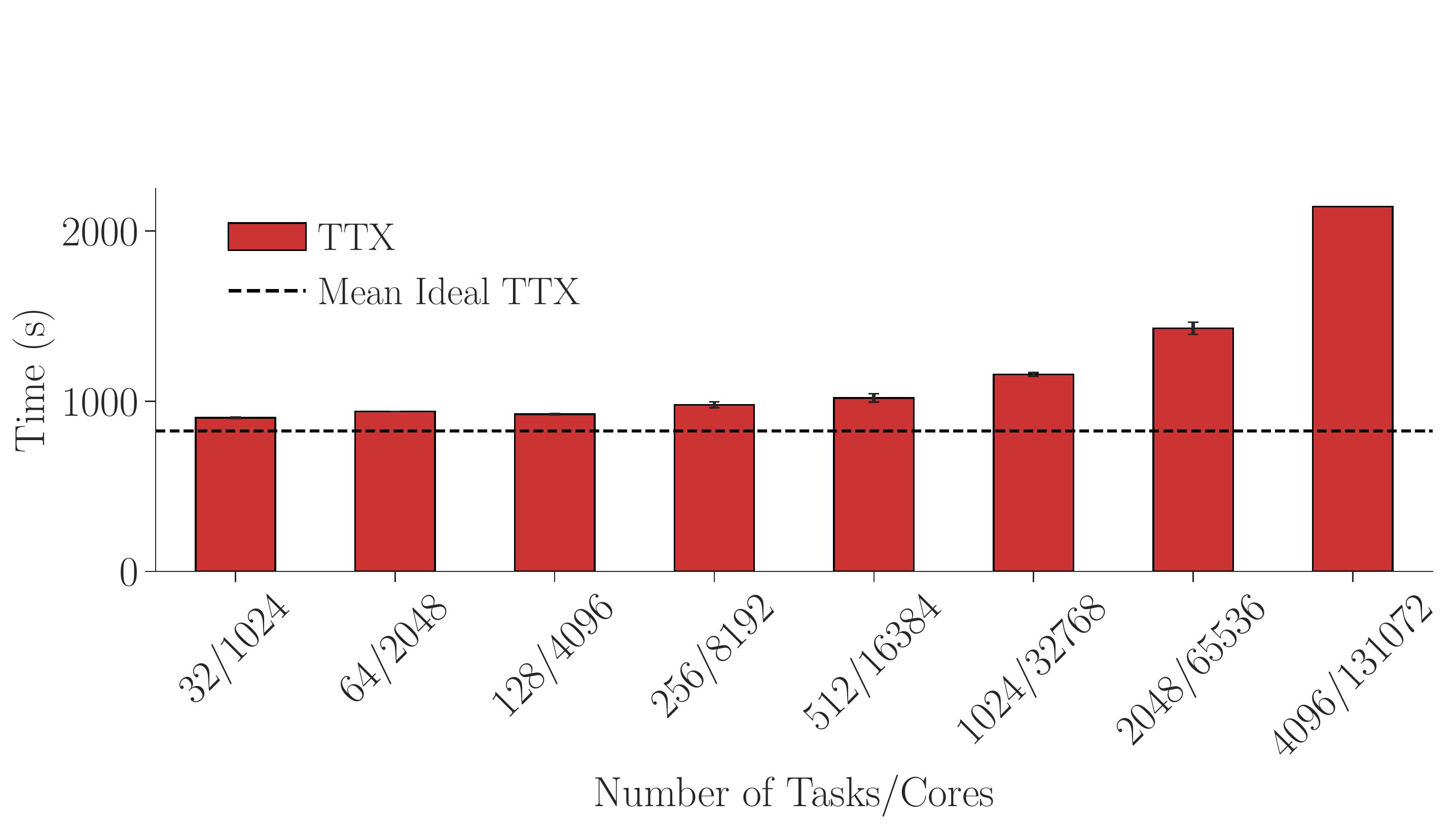}
  \includegraphics[width=0.49\textwidth,valign=t]{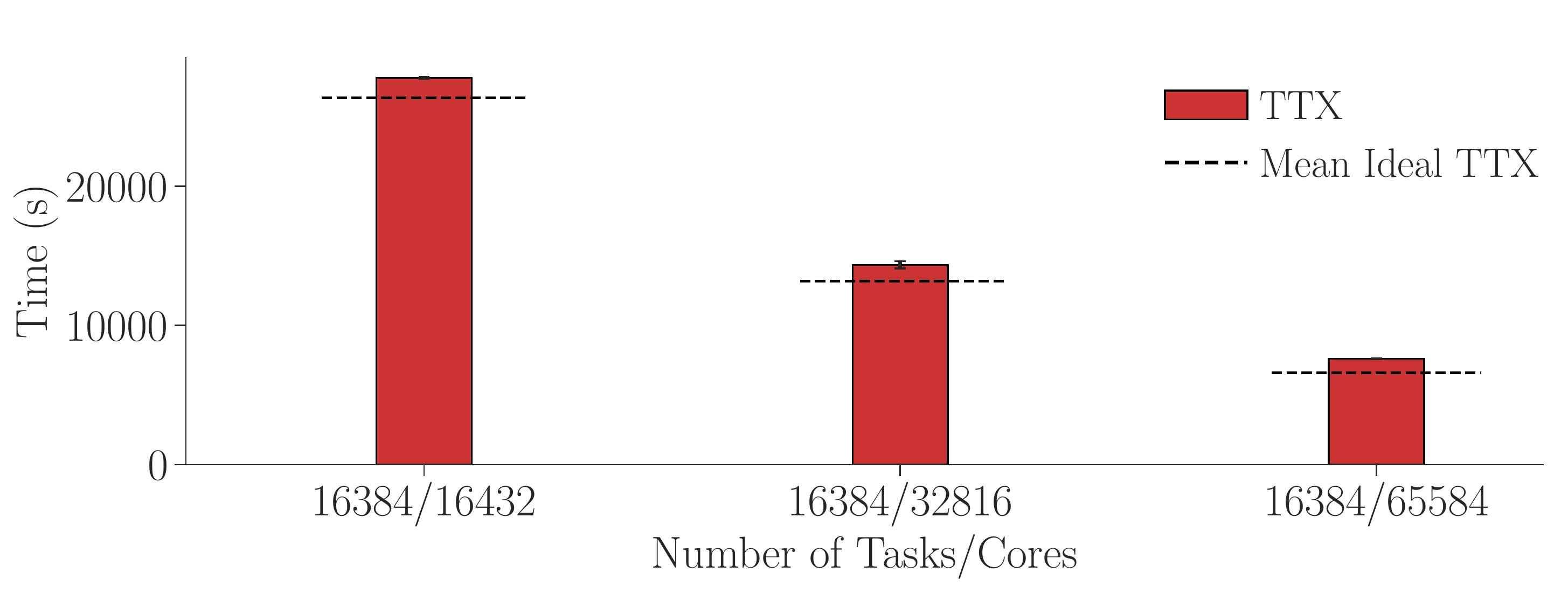}
  \up\up
  \caption{\textbf{Experiments 1--2:} RP weak (top) and strong (bottom)
    scaling.\label{fig:s-ttc}}
  \up\up
\end{figure}

Fig.~\ref{fig:s-ttc} shows the scaling of RP for the workloads of experiments 1
and 2 (Table~\ref{tab:experiments}). An ideal TTX (broken line) represents
execution time without RP and resource overheads, and corresponds to the mean
value in Fig.~\ref{fig:tx_distribution}. In experiment 1, the ratio between
number of tasks and core is constant, enabling fully concurrent executions.

Fig.~\ref{fig:s-ttc}~(top) shows that the actual TTX scales almost linearly
between 1024 and 4097 cores, and sublinearly between 4097 and 131,072 cores. The
average value of TTX for runs with between 1024 and 4097 cores is 922\(\pm\)14
seconds (s), indicating an average overhead of 11\% over the mean of the ideal
TTX\@. This overhead grows between 18\%/160\% at 8192/131,072 cores.

Fig.~\ref{fig:s-ttc}~(bottom) shows the strong scaling of 16,384 tasks
executed from 16,384 to 65,536 cores; this results in the number of generations
varying from 32 to 8. When executed over 16,384, 32,816 and 65,536 cores, they
have a TTX of 27,794s\(\pm\)70, 14,358s\(\pm\)259, and 7612s\(\pm\)29
respectively. The deviation from ideal TTX is relatively uniform across
different pilot sizes (1,158s\(\pm\)150), which indicates that RP is less
efficient at higher pilot core counts.

\begin{figure}
  \centering
  \includegraphics[trim=0 0 0 25,width=0.49\textwidth]{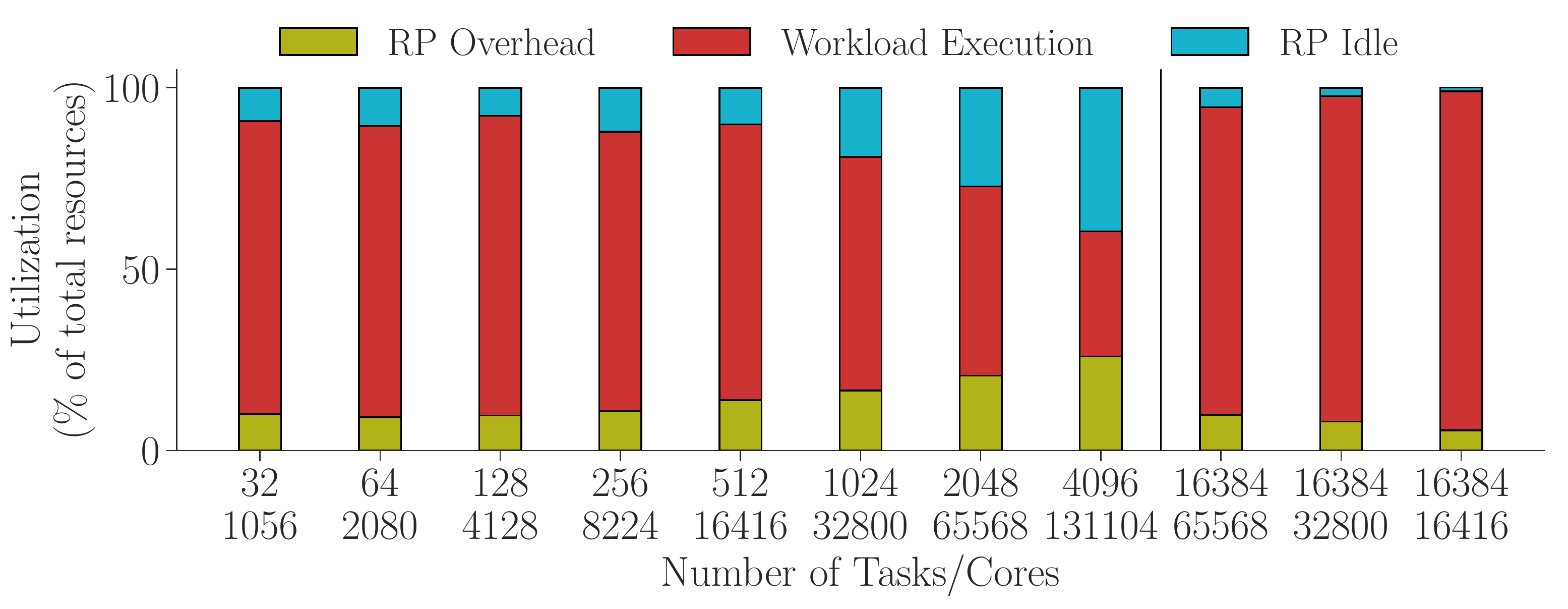}
  \up\up\up
  \caption{\textbf{Experiments 1 and 2:} Resource utilization (RU) of
  RADICAL-Pilot. First 8 bars: experiment 1; Last 3 bars: experiment
    2.\label{fig:ws-ru}}
  \up\up
\end{figure}

Fig.~\ref{fig:ws-ru} shows RU for experiment 1 (first 8 bars) and experiment 2
(last 3 bars). RU is represented as the percentage of the available core-time
spent executing the workload, RP components, third party software (i.e., ORTE,
the lunch method used on Titan to execute tasks) or idling. Note the relation
between TTX and RU\@: The more core-time is spent executing the workload, the
shorter TTX\@.

Fig.~\ref{fig:ws-ru}~(first 8 bars) shows for experiment 1 a relatively constant
percentage of core-time utilization for runs with between 32--128 tasks and
1024--4097 cores, consistent with TTX of Fig.~\ref{fig:s-ttc}~(top). The
percentage of utilization decreases with the growing of the number of
tasks/cores, also consistent with Fig.~\ref{fig:s-ttc}~(top). Interestingly, our
analysis of the traces shows that there are three main reasons for the
decreasing of resource utilization: scheduling, ORTE and idling.

For experiment 2, Fig.~\ref{fig:ws-ru}~(last 3 bars) shows progressively shorter
values for RP scheduling, ORTE and idling for runs with multiple generations (as
defined in~\S\ref{ssec:exp_design}). When tasks of one generation terminate,
those of the following generation immediately start executing. This eliminates
the idling of cores for all generations but the last. Further, RP and ORTE
overheads increase with the number of cores, indicating that the reduced
performance of RP measured in Fig.~\ref{fig:s-ttc}~(top) depends on the size of
the pilot. Note that the more generations in a strong scaling run, the
longer the runtime and that, the longer the runtime, the less relevant RP
Overhead and RP Idle become for the percentage of overall resource utilization.

Together, the data of experiments 1 and 2 show that the challenges of scaling
homogeneous task execution beyond 4097 cores mostly depends on the efficiency of
RP's scheduler and of ORTE's launching system at managing concurrent executions
on pilot of growing size.

\up\up
\subsection{Improving Performance and Scale}\label{ssec:exp_agent}

\begin{figure*}
  \centering
  \includegraphics[trim=0 0 0 80,clip,width=\textwidth]{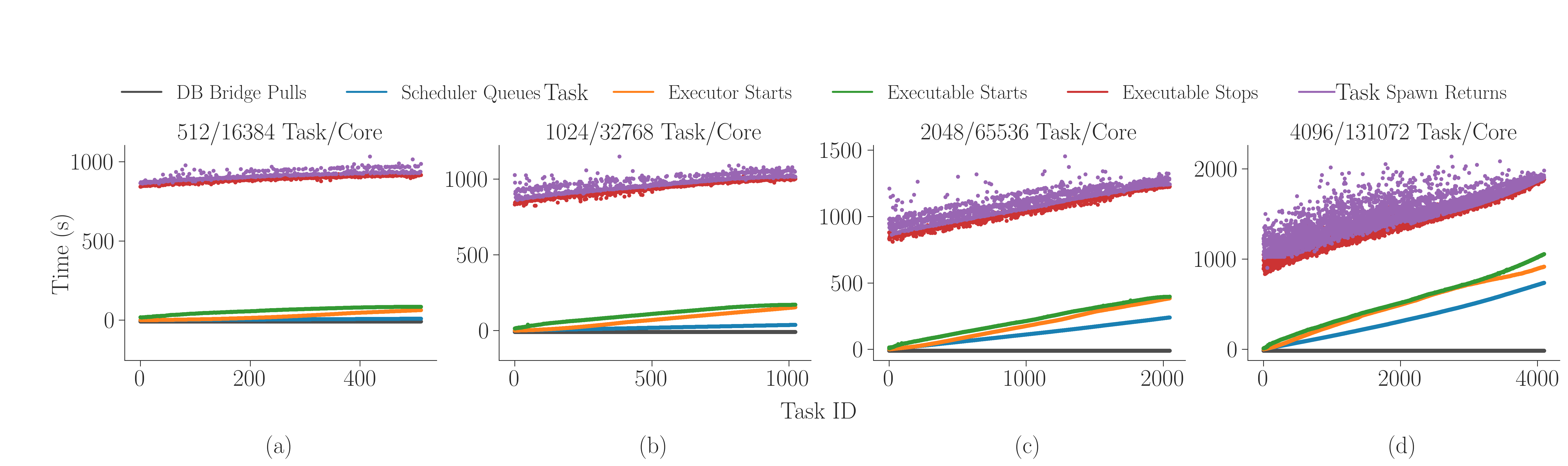}
  \up\up\up\up
  \caption{\textbf{Experiments 1:} Tasks events. Scheduler (blue) and Executor
  (purple) events limit the weak scaling of RP. Different Y axes to
    improve readability.\label{fig:ws-events}}
  \up\up\up
\end{figure*}

Fig.~\ref{fig:ws-events} clarifies the relation between the performance of the
Scheduler and the Executor, the two Agent components that, alongside ORTE,
contribute to RP's overhead in experiments 1 and 2. We measure the time spent by
each task in each component of the Agent. Tasks are pulled from RP DB into the
Scheduler's queue (Fig.~\ref{fig:ws-events}, DB Bridge Pulls, black); after, the
Scheduler queues each task into an Executor (Fig.~\ref{fig:ws-events}, Scheduler
Queues Task, blue); the Executor starts processing the queued task
(Fig.~\ref{fig:ws-events}, Executor Starts, orange), starting task's executable
(Fig.~\ref{fig:ws-events}, Executable Starts, green) and waiting for it to stop
(Fig.~\ref{fig:ws-events}, Executable Stops, red) executing. Finally, the
Executor marks the task as done (Fig.~\ref{fig:ws-events}, Task Spawn Returns,
purple).

Fig.~\ref{fig:ws-events} shows that all the tasks of the workload, pulled in
bulk from the DB (DB Bridge Pulls), enter Scheduler's queue approximately at the
same time; i.e., all the tasks are approximately at the same height compared to
the y-axes, forming an almost horizontal ``line'', parallel to the x-axis. The
angle between the black line (DB Bridge Pull) and the blue line (Scheduler
Queues Task) is a measure of the time taken by RP to schedule each task. The
wider the angle, the more time scheduling takes. Ideally, tasks should be
immediately scheduled for execution as in experiment 1 there are as many cores
available as needed by all the tasks.

Fig.~\ref{fig:ws-events} also shows two overheads in Executor that depend on
ORTE and not RP\@: (1) the time spent to prepare a task for its execution
(Executor Starts), i.e., the time between when a task is passed to ORTE and when
it starts to execute; and (2) the time required for the Executor to be informed
that a task has been executed (Task Spawns Return), i.e., the time from when a
task stops executing and the time when ORTE passes a message to the Executor
about the task being done or failed. The mean time to prepare the execution of
512 tasks on 16,384 cores is 37s\(\pm\)9; 37s\(\pm\)6 with
1024~tasks/32,768~cores; 35s\(\pm\)8 with 2048~tasks/65,536~cores; and
41s\(\pm\)30 with 4096~tasks/131,072~cores. Thus, in spite of the high jitter,
the mean is essentially invariant across scales.

The Executor takes variable amount of time to acknowledge that the execution of
a task has completed. This variance increases with scale, depending on the time
taken by ORTE to communicate with RP about the task's state and the time taken
to process the message. The distribution of the Task Spawn Returns event is both
broad and long-tailed across all the scales. The mean time to communicate the
completion of 512 tasks on 16,384 cores is 29s\(\pm\)16; 34s\(\pm\)28 with
1024~tasks/32,768 cores; 59s\(\pm\)46 with 2048~tasks/65,536~cores; and
135s\(\pm\)107 with 4096~tasks/131,072~cores.

Based on that analysis, we improved RP performance by implementing a more
efficient scheduling algorithm, using PRRTE instead of ORTE and reducing the
time spent idling while resources are available to execute tasks. Experiments 3
and 4 measure the improved performance at scale of RP and execute heterogeneous
workloads on heterogeneous resources, moving away from the homogeneity of
experiments 1 and 2. Note that, due to the workload, platform, RP scheduler and
RP executor, the results of experiments 1 and 2, and 3 and 4 are not directly
comparable.

\up\up
\subsection{Experiments 3--4: Weak and Strong Scaling with Heterogeneous Tasks
on Heterogeneous Resources}\label{ssec:exp_3_4}

\begin{figure*}
   \centering
   \includegraphics[width=\textwidth]{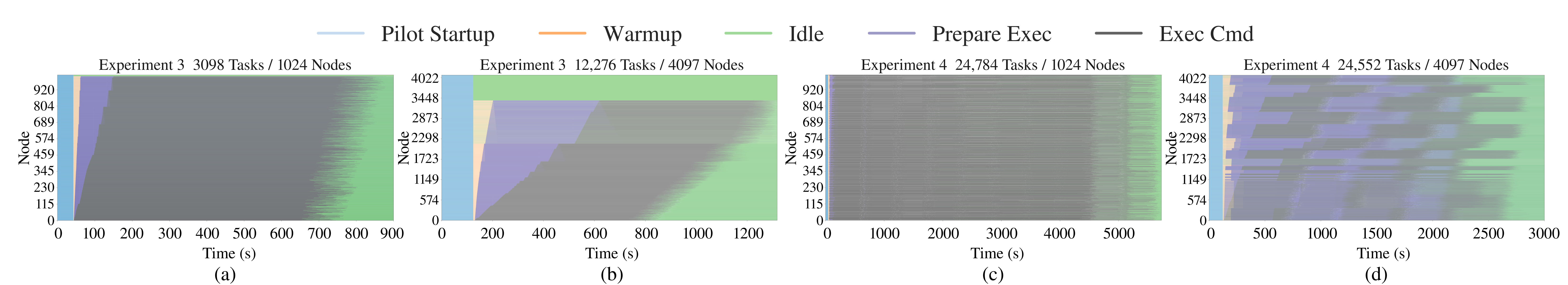}
    \up\up\up
   \caption{\textbf{Experiments 3--4:} Weak (a,b) and strong (c,d) scaling of
   RP resource utilization when executing 3098, 12,276, 24,784 and 24,552 tasks
   on 1024 and 4097 Summit compute nodes. Tasks are heterogeneous for duration,
   number of CPUs/GPUs, number of threads/processes, and use of
    MPI.\label{fig:summit-ws-ss}}
    \up\up
  \end{figure*}

Fig.~\ref{fig:summit-ws-ss} shows RP resource utilization (RU) for experiments 3
and 4. \textit{Pilot Startup} (blue) shows the time in which the resources are
blocked while RP starts up; and \textit{Warmup} (orange) the time in which
resources are blocked by RP while collecting tasks and scheduling them for
execution. \textit{Prepare Exec} (purple) indicates the resources blocked while
waiting for PRRTE to initiate task execution; \textit{Exec Cmd} (black) marks
the time in which tasks use resources for execution; and \textit{Idle} (green)
the time in which available resourced idled.

Compared to experiments 1 and 2, we improved the scheduler performance from $6$
to $300$ tasks/s, eliminated the delay in the acknowledgment of task completion
by using PRRTE instead of the now deprecated ORTE, and partitioned the execution
across multiple DVMs. As a result, RP scheduled 3098 tasks on 1024 compute nodes
(43,008 cores/6144 GPUs) in $\sim$10s (Fig.~\ref{fig:summit-ws-ss}a, yellow area)
and 12,276 tasks on 4097 compute nodes (172,074 cores/24,582 GPUs) in
$\sim$100s (Fig.~\ref{fig:summit-ws-ss}b, yellow area), achieving linear
scaling performance. Both runs used PRRTE with up to 256 nodes per DVM, thus 4
DVMs for 1024 nodes and 16 DVMs for 4097 nodes with 1 node reserved to RP Agent.
In this configuration, we measured a negligible overhead for acknowledging task
completion and thus addressed the performance issue measured with ORTE. Note
that in the second run of experiment 3, two DVMs failed
(Fig.~\ref{fig:summit-ws-ss}b, unused resources on the top) but, due to RP
fault-tolerance, all the tasks were executed on the remaining DVMs.

Figs.~\ref{fig:summit-ws-ss}a and~b show that, once RP Executor has passed the
tasks to PRRTE, the time PRRTE takes to launch those tasks increases with the
number of the available resources (purple area). Based on
Ref.~\cite{turilli2019characterizing}, we know that PRRTE and DVM overheads are
relatively small when managing up to 16,000 tasks on up to 400 Summit compute
nodes. Our analysis confirmed that the observed performance degradation depends
on the performance of the file system. When executing at full capacity, the
distributed filesystem on which PRRTE is installed shows that it was not
designed and optimized for large amounts of (relatively) small concurrent I/O.
This problem might be mitigated by installing PRRTE on each compute node when
bootstrapping RP but that would affect both overheads and resource utilization.

Experiment 3 runs reached 77\% and 41\% resource utilization with 3098/12,276
tasks and 1024/4097 nodes respectively. The lower utilization of the run with
4097 nodes is due to the file system overheads described above: the delayed
starting of task execution wastes resource availability but also increases the
time spent waiting for those tasks to complete (Fig.~\ref{fig:summit-ws-ss}b,
green area). As a consequence of how HPC resource managers work, RP has to wait
for all the tasks to complete before releasing all the acquired resources.
Another $\sim$10\% of utilization is lost due to the failure of 1148 tasks. That
is mostly due to PRRTE mishandling processes under the pressure of concurrency,
something that needs to be improved in PRRTE/PMIx.

Figs.~\ref{fig:summit-ws-ss}c and~\ref{fig:summit-ws-ss}d confirm that improved
scheduling rate and reduced PRRTE task acknowledging time hold also with the
strong scaling runs of experiment 4. RP reached 76\% resource utilization with
24,784 tasks / 1024~nodes and 38\% with 24,552~tasks / 4097~nodes. The filesystem
issues already observed in experiment 3 multiply in experiment 4 due to the
presence of multiple generations (Fig.~\ref{fig:summit-ws-ss}d, multiple purple
areas) and compound to the overheads of managing workload heterogeneity over
multiple generations, affecting the overall resource utilization. RP scheduler
could use better bin packing algorithms but the best results would require
accurate task duration estimation which is difficult to obtain in production
scenarios. Currently, the best approach would be to use RP multi-pilot
capabilities to partition the workload across 4 independent pilots and benefit
from the better performance measured with 1024 nodes.

RP overhead (OVH) for experiments 3 and 4 is: 61s (3098~tasks / 1024~nodes), 131s
(12,276~tasks / 4097~nodes), 115s (24,784~tasks / 1024~nodes), and 251s
(24,552~tasks / 4097~nodes). Barring the scheduling overhead (yellow areas in
Fig.~\ref{fig:summit-ws-ss}) most of the overhead is due to the time taken to
bootstrap the agent (blue areas in Fig.~\ref{fig:summit-ws-ss}). Bootstrap
overhead is invariant to walltime and thus it becomes less relevant for
production-grade workloads that usually run for many hours. In
Fig.~\ref{fig:summit-ws-ss}d, PRRTE took more time than usual to tear down the
DVMs (green area), increasing the OVH of that run.

Overall, the performance and scalability limits outlined by experiments 3 and 4
are those of PRRTE/PMIX which we use as system execution layer. RP itself
behaves as expected: it timely schedules tasks and passes them on to the
execution layer. It should also be noted that Summit's native execution layer
(\texttt{LSF}/\texttt{jsrun}) has much lower scalability limits of about 800
concurrent tasks~\cite{turilli2019characterizing}.

Resources partitioning is the way forward to improve the performance of RP on
the upcoming exascale platforms, while reducing the impact of other systems'
overheads as experienced with PRRTE. We will partition RP Agent, add a
Metascheduler component and deploy a Scheduler and Executor for each partition.
The size and lifespan of each partition will be dynamic, allowing to minimize
the amount of resources assigned to each partition, based on the requirements of
the tasks that will execute on those resources. Barring workloads with unusually
large MPI tasks and given the current capabilities of HPC platforms, the
aggregated performance of all the partitions will be higher than that of a
single, machine-wide partition. That is the approach we started to explore with
multiple DVMs and multiple masters/workers in experiment 5.

\up\up
\subsection{Experiment 5: Function Calls Execution on Multiple Pilots}
\label{ssec:exp_5}

\begin{figure*}
   \up\up
   \centering
   \subfloat[][]{
      \includegraphics[width=0.31\textwidth]{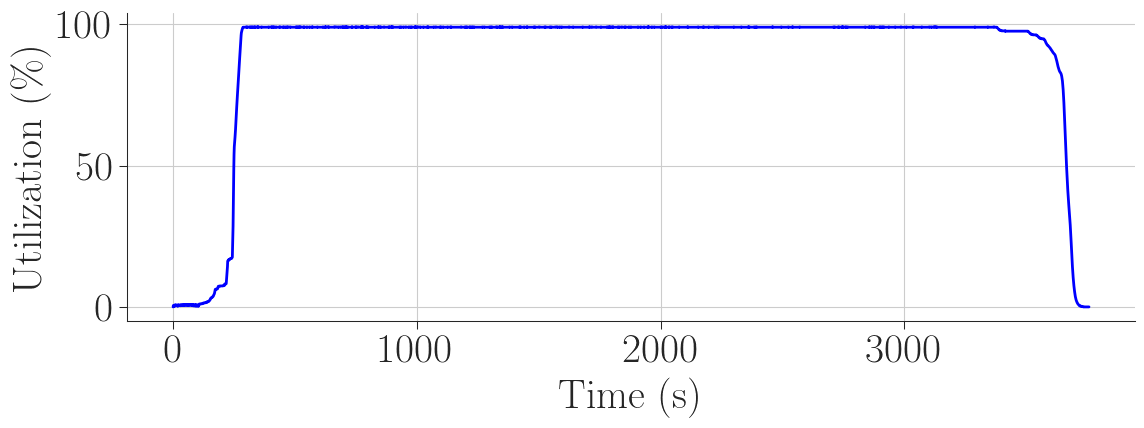}
      \label{sfig:frontera-ru}}
   \hfill
   \subfloat[][]{
      \includegraphics[width=0.31\textwidth]{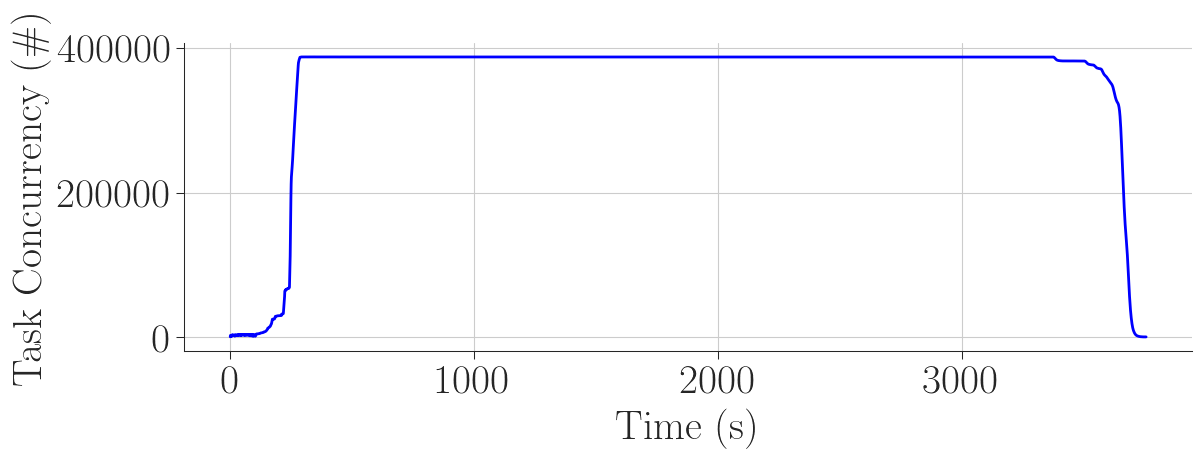}
      \label{sfig:frontera-tc}}
   \hfill
   \subfloat[][]{
      \includegraphics[width=0.31\textwidth]{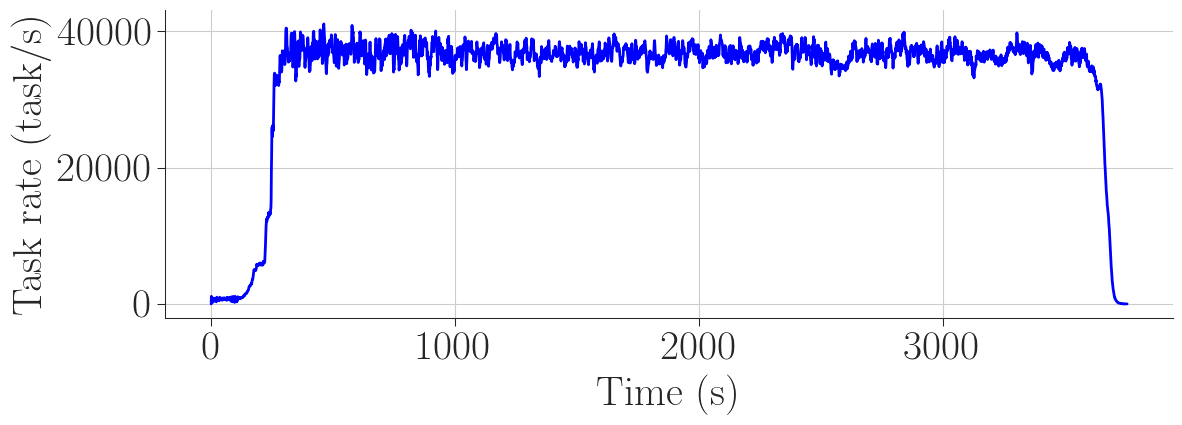}
      \label{sfig:frontera-tr}}
   \up
   \caption{\textbf{Experiments 5:} RP (a) resource utilization (RU), (b)
   execution concurrency (EC) and (c) task execution rate (TR) with RAPTOR when
   executing 126,471,524 OpenEye Python function calls on 7000 compute
   nodes/392,000 cores of Frontera with 70 master and 99 workers per master. RU
   = 90\%; EC = $4\times10^5$ steady state; TR = $144\times10^6/hour$
    peak.\label{fig:frontera-raptor}}
   \up\up\up
  \end{figure*}

Fig.~\ref{fig:frontera-raptor} shows resource utilization
(Fig.~\ref{sfig:frontera-ru}), task execution concurrency
(Fig.~\ref{sfig:frontera-tc}) and task execution rate
(Fig.~\ref{sfig:frontera-tr}) over the time taken by RP and RAPTOR to execute
the 126,471,524 OpenEye Python function calls of experiment 5. Partitioning the
resources across 70 masters, each managing 99 workers, RP and RAPTOR utilize
90\% of the available resources, reaching 98\% utilization after $\sim$300s and
keeping that rate for $\sim$3000s, i.e., 80\% of the overall runtime. RP takes
less than 300s to bootstrap and to launch the 70 masters and 6930 workers. The
tapering down of the resource utilization towards the end of the execution
depends on the differences in each of the data processed by the function call
(i.e., the physical properties of the receptor that is docked) and on the
progressive exhaustion of the calls that still need to be executed.

Figs.~\ref{sfig:frontera-tc} and~\ref{sfig:frontera-tr} are consistent with the
resource utilization plotted in Fig.~\ref{sfig:frontera-ru}. After initial
warm up, RP and RAPTOR reach steady state, executing $\sim$390,000
concurrent tasks/s at every point in time until the 3000s mark of the
runtime, saturating the available 392,000 cores. Task execution rate indicates
the number of tasks completed over time and Fig.~\ref{sfig:frontera-tr} shows
that it averages 37,000 tasks/s with peaks of 40,000 tasks/s. This is consistent
with the concurrency rate, the average task execution time of 34s, the
number of cores concurrently available and the total number of tasks to compute.

Experiment 5 and the use of multiple masters and workers confirms what already
observed with experiments 3 and 4: partitioning of resources is a promising
approach to limit global overheads, while improving resource utilization within
each partition. Further, our experiments on Frontera showed the importance of
tailoring the HPC platform capabilities to the requirements of many-tasks
workflows. TACC system administrators configured one of the shared filesystem so
to better support the load of our type of workload, and tailored libraries and
Python to reduce I/O to a minimum.

\up\up
\section{Conclusions}\label{sec:conclusion}

Software systems implementing the Pilot
abstraction~\cite{turilli2018comprehensive} provide the conceptual and
functional capabilities to support the scalable execution of workloads comprised
of many heterogeneous tasks. Whereas there are multiple Pilot systems, they are
geared towards either specific workloads or platforms. Against this backdrop,
RADICAL-Pilot (RP) brings together conceptual
advances~\cite{turilli2018comprehensive} with systems and software
engineering~\cite{turilli2019middleware} showing potential for portability,
extendibility and performance at extreme scale.

This paper describes RP's design and implementation (\S\ref{sec:arch}), and
characterizes the performance of its Agent module on past and present HPC
leadership-class machines for homogeneous, heterogeneous and production
workloads (\S\ref{sec:exp}). Although RP works on multiple platforms, we focused
our experiments on existing leadership-class platforms that offer the highest
degree of concurrency both for CPUs cores and GPUs, and that are precursors to
the first generation of exascale platforms. The experiments discussed
in~\S\ref{sec:exp} benefited from RP's support for tracing and profiling. Using
RADICAL-Analytics, we were able to pinpoint and reduce RP overheads while
isolating performance bottleneck of the HPC platform and third-party software
tools.

Experiments 1 and 2 in~\S\ref{sec:exp} outlined the relevant scheduling
performance, the limitations of launching systems and, ultimately, indicated the
need to partition resources at different logical levels. Experiment 3 and 4
showed that by addressing those limitations, we were able to scale workload
executions on the largest HPC platform with heterogeneous compute resources.
Further, experiments 3 and 4 also showed how RP can manage multiple dimensions
of heterogeneity at large scales, without incurring limiting overheads. Finally,
experiment 5 showed how RP can be effectively and efficiently used to execute
hundred of millions of Python function calls on NSF Frontera. In fact, RP
enabled approximately $150\times10^6$ docks/hour, about two times the highest
known published rate~\cite{vermaas2020supercomputing}.

The focus of this paper has been on the direct execution of workloads on HPC
machines, but RP also forms the middleware and runtime system for a range of
other tools and libraries, already used in production. RP was designed following
the `building blocks approach', enabling integration with third-party software
systems such as Parsl, Swift, PanDA and Flux. RP is available for immediate use
on many HPC platforms~\cite{radical_pilot_url}, accompanied with documentation
and an active developer-user community.

This paper offers several indications of what is needed to enable the execution
of heterogeneous workloads on the upcoming exascale HPC platforms. Partitioning
executions across multiple third-party launchers (e.g., DVMs) proven to be
effective but limited due to the overheads posed by load balancing among
different launchers. We plan to implement multiple levels of partitioning at the
Agent, Scheduler and Executor level. In this way, we will benefit from
multi-stage placement, not only distributing the overheads across different
subsystems but also decoupling, as much as possible, the magnitude of the
overheads from the scale of the concurrency at which the workload will be
executed. Further, this approach will also improve error handling, fault
tolerance and resilience.

Another important message of this paper is the need for considering
heterogeneous workloads, and thus workflows, as a first-order priority of the
exascale roadmap. As pointed out in the introduction, such workflows are
becoming ubiquitous in many scientific domains and the demand for scale and
performance had reached critical mass. The performance limits of Summit's file
system measured in~\S\ref{sec:exp}, Experiment 3, underline the importance of
considering the requirements of heterogeneous, many-task workflows when
designing the upcoming exascale machines.

This paper also shows the importance of producing a benchmark suite for pilot
systems and HPC platforms. Currently, it is difficult if not impossible to
compare RP performance to other pilot systems because the lack of common
metrics, analogous task implementations, and effective ways to isolate a
platform, pilot system and task overheads. Further, a benchmark suite would also
be necessary to validate the effectiveness of future HPC platforms in supporting
diverse workflows. Proposed performance enhancements of RP will benefit from
such benchmarks, while being the runtime system of workflow benchmarks used in
the procurement of future leadership platforms.

\noindent {\bf Acknowledgements} {\footnotesize This research is supported by
NSF RADICAL-Cybertools (NSF-1931512 ) and SCALE-MS (NSF-1835449), ECP CANDLE and
ExaWorks, and the Exascale Computing Project (17-SC-20-SC) at BNL (contract
DESC0012704). This research used OLCF resources at ORNL (contract
DE-AC05-00OR22725), and NSF XSEDE resources (allocation TG-MCB090174). We thank
TACC for the opportunity for scaling runs during the TexaScale days. We thank
Mark Santcroos and Manuel Maldonado for early stage contributions.}

\noindent {{\bf Experiments} Data and analysis scripts
can be found at: {\footnotesize \url{https://github.com/radical-experiments/rp.paper}}

\small

\bibliographystyle{IEEEtran}
\bibliography{rp}

\begin{IEEEbiography}
[{\includegraphics[trim=20 0 20 0,clip,width=1in]{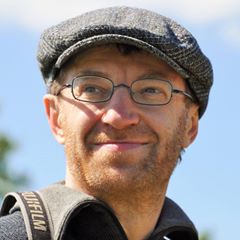}}]{
    Andre Merzky}
Andre Merzky received his diploma in Physics in 1998 at the Humboldt
University in Berlin. He has worked since on Grid-related topics concerning
data management, visualization, distributed and high performance computing.
He is architect and lead developer of RADICAL-Pilot.
\vspace{-15 mm}
\end{IEEEbiography}

\begin{IEEEbiography}
[{\includegraphics[trim=0 0 0 0,clip,width=1in]{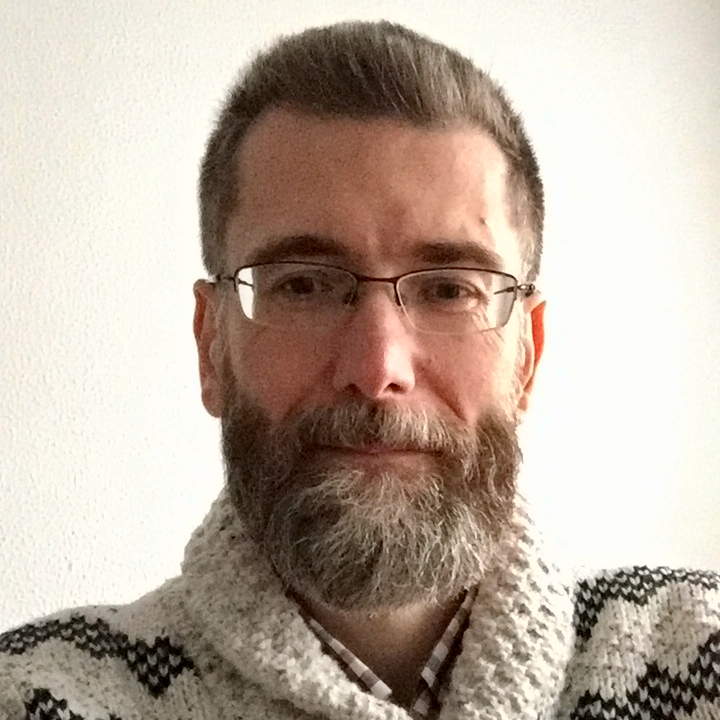}}]{
   Matteo Turilli}
Matteo Turilli is Assistant Research Professor at Rutgers University, Electrical
\& Computer Engineering Department. His research focuses on merging
high-throughput and high-performance computing, enabling the execution of
heterogeneous workflows on among the largest computing platforms in the world.
He holds a DPhil in Computer Science from the University of Oxford, UK.
\vspace{-15 mm}
\end{IEEEbiography}

\begin{IEEEbiography}
[{\includegraphics[trim=100 0 200 0,clip,width=1in]{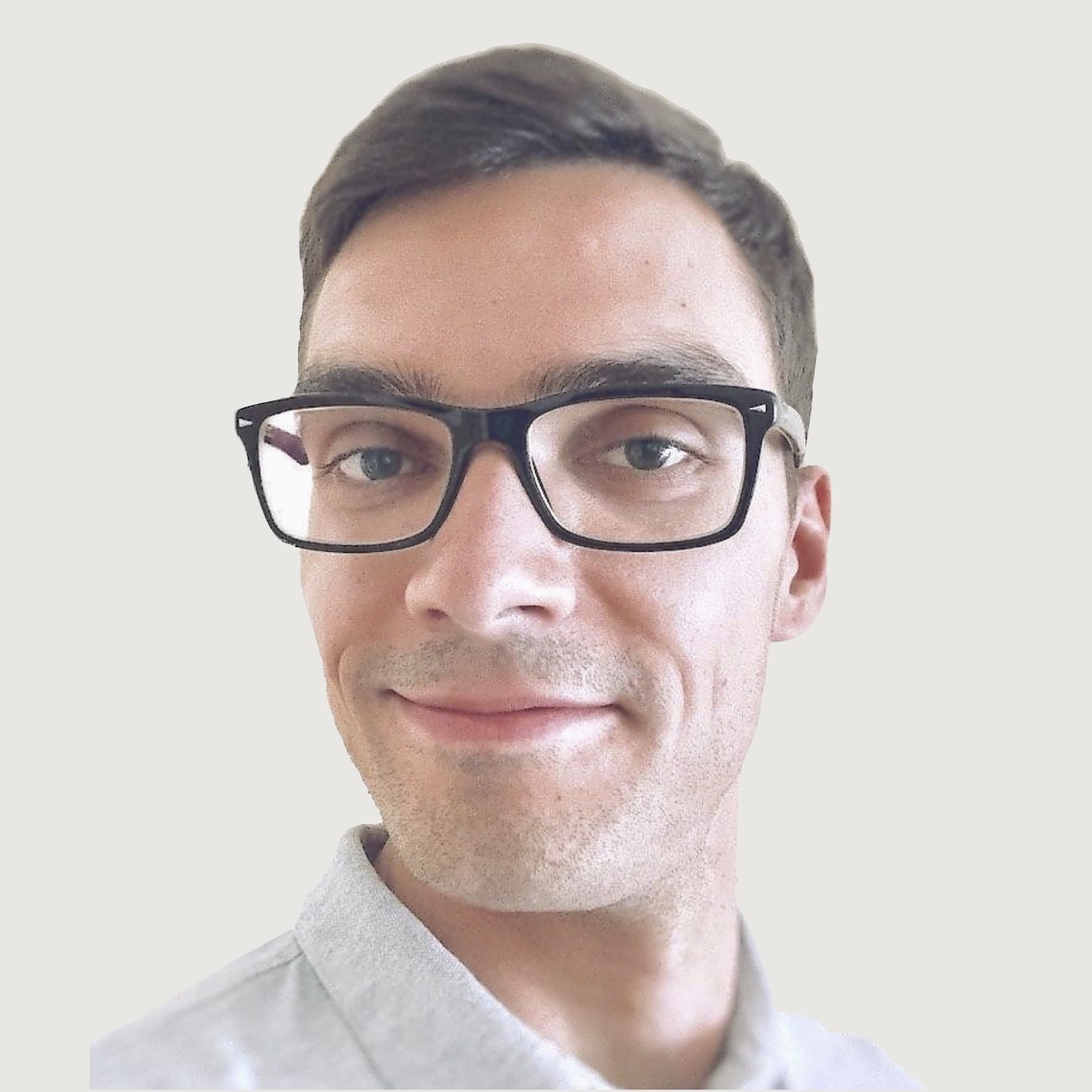}}]{
   Mikhail Titov}
Mikhail Titov is a Research Associate at Rutgers University. He pursued his
PhD degree in Computer Science at the University of Texas at Arlington in 2016.
Since 2008, he is an Associated member of the personnel (Scientist) at
the European Organization for Nuclear Research (CERN, Geneva, Switzerland).
His research areas are grid computing, cloud computing, high performance
computing, data mining, machine learning, modeling and simulation.\vspace{-10 mm}
\vspace{-5 mm}
\end{IEEEbiography}

\begin{IEEEbiography}
[{\includegraphics[trim=840 2230 1010 1220,clip,width=1in]{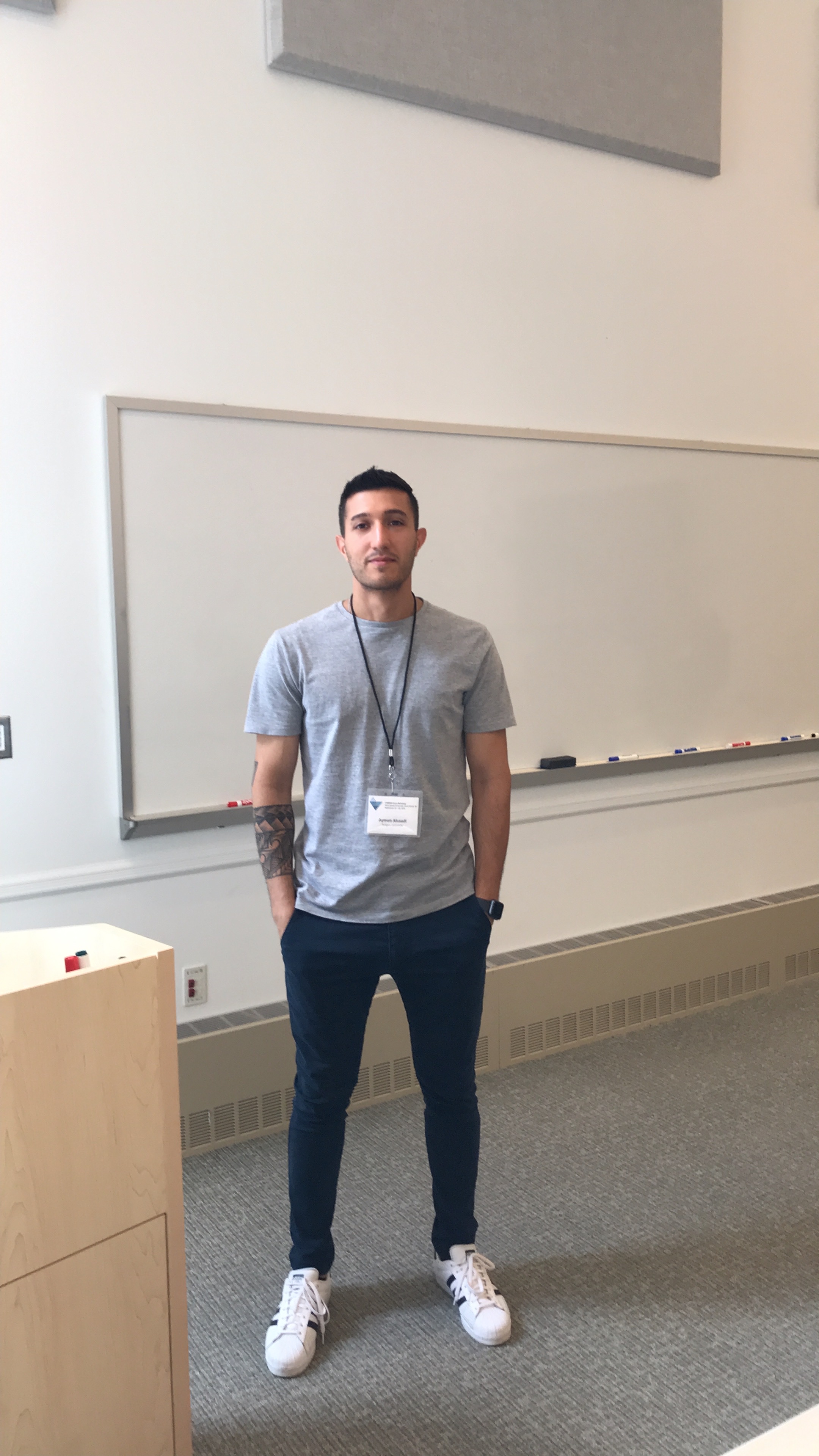}}]{
   Aymen Al-Saadi}
Aymen Al-Saadi is a PhD student and Junior Research Developer of the RADICAL
group. Before that he earned his master degree in Computer Engineering from
Rutgers University.  His research focus currently lies on high throughput
function execution in HPC contexts.\vspace{-10 mm}
\vspace{-5 mm}
\end{IEEEbiography}

\begin{IEEEbiography}
[{\includegraphics[trim=80 0 30 0,clip,width=1in]{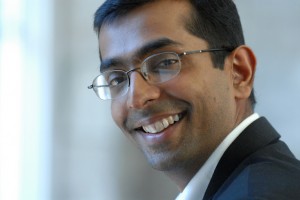}}]{
   Shantenu Jha}
Shantenu Jha is a Professor of Computer Engineering at
Rutgers University and the Chair of the Department (Center) for Data Driven
Discovery at Brookhaven National Laboratory. He was appointed a Rutgers
Chancellor’s Scholar in 2015. Shantenu’s research interests are at the
intersection of high-performance distributed computing and computational
\& data-driven science.
\vspace{-5 mm}
\end{IEEEbiography}

\end{document}